\documentclass[aps,preprint,floatfix,nofootinbib,showpacs]{revtex4-1}
\pdfoutput=1
\usepackage{graphicx,color,float}
\usepackage{hyperref}

\begin{document}

{\small
\begin{flushright}
CNU-HEP-16-01
\end{flushright} }

\title{
Entangling Higgs production associated with a single top
and a top-quark pair in the presence of
anomalous top-Yukawa coupling
}

\def\slash#1{#1\!\!/}
\def\lsim{\:\raisebox{-0.5ex}{$\stackrel{\textstyle<}{\sim}$}\:}
\def\gsim{\:\raisebox{-0.5ex}{$\stackrel{\textstyle>}{\sim}$}\:}

\renewcommand{\thefootnote}{\arabic{footnote}}

\author{
Jung Chang$^1$, Kingman Cheung$^{1,2,3}$, Jae Sik Lee$^{4,1}$, and
Chih-Ting Lu$^3$}
\affiliation{
$^1$ Physics Division, National Center for Theoretical Sciences,
Hsinchu, Taiwan \\
$^2$ Division of Quantum Phases and Devices, School of Physics, 
Konkuk University, Seoul 143-701, Republic of Korea \\
$^3$ Department of Physics, National Tsing Hua University,
Hsinchu 300, Taiwan \\
$^4$ Department of Physics, Chonnam National University, \\
300 Yongbong-dong, Buk-gu, Gwangju, 500-757, Republic of Korea
}
%\pacs{14.80.Bn.,14.80.Da,14.80.Ec}
%\date{\color{red}\today}
%\date{July 22, 2016}
%\date{November 1, 2016}
%\date{January 3, 2017}
%\date{February 22, 2017}
\date{April 10, 2017}

\begin{abstract}
The ATLAS and CMS collaborations observed
a mild excess in the associated Higgs production with a
top-quark pair ($t\bar t h$)  
and reported the signal strengths of 
$ \mu_{tth}^{\rm ATLAS}=1.81\pm 0.80$
and 
$\mu_{tth}^{\rm CMS}=2.75\pm 0.99 $
based on the data collected at $\sqrt{s}$= 7 and 8 TeV.
Although, at the current stage,
there is no obvious indication whether the excess is 
real or due to statistical fluctuations,
here we perform a case study of this mild
excess by exploiting the strong entanglement between the associated Higgs 
production with a single top quark ($thX$)  and $t\bar t h$ production
in the presence of anomalous top-Yukawa coupling.
As well known,
$t\bar t h$ production only depends on the absolute value of the
top-Yukawa coupling.  Meanwhile, in $thX$ production,
this degeneracy is lifted through the strong
interference between the two main contributions which are proportional to the
top-Yukawa and the gauge-Higgs couplings, respectively.
Especially, when the relative sign of the top-Yukawa coupling
with respect to the gauge-Higgs coupling
is reversed, the $thX$ cross section can be
enhanced by more than one order of magnitude.
We perform a detailed study of the influence of $thX$ 
production on $t\bar{t}h $ production
in the presence of the anomalous top-Yukawa coupling
and point out that
it is crucial to include $thX$ production in the analysis of 
the $t\bar t h$ data to pin 
down the sign and the size of the top-Yukawa coupling in future.
While assuming the Standard Model (SM) value for the gauge-Higgs coupling,
we vary the top-Yukawa coupling within the range allowed by the current 
LHC Higgs data.
We consider the Higgs decay modes into multileptons, 
$b\bar b$ and $\gamma\gamma$ putting a particular emphasis 
on  the same sign dilepton events.
We also discuss the
prospects for the LHC Run-2 on how to disentangle $thX$ 
production from $t\bar{t}h$ one and how to
probe the anomalous top-Yukawa coupling.
\end{abstract}

\maketitle

\section{Introduction}
The Higgs boson was discovered at the Large Hadron Collider (LHC)
\cite{atlas,cms}. After analyzing almost all the Run-1 data, the
measured properties of the Higgs boson are the best described by the
standard model (SM) Higgs boson \cite{higgcision}, which was proposed
in 1960s \cite{higgs}. The most constrained is 
the Higgs coupling to the massive gauge bosons 
normalized to the corresponding SM value
(gauge-Higgs couplings) $C_v = 0.94\,^{+0.11}_{-0.12}$,
which is very
close to the SM value \cite{Khachatryan:2014jba}. 
On the other hand, the top- and
bottom-Yukawa couplings cannot be determined as precisely as $C_{v}$ by
the current data. Currently, they are within $ 30-40\% $ of the SM
values \cite{Khachatryan:2014jba}, yet, 
the negative regime of the top-Yukawa coupling
is still allowed at 95\% confidence level (CL)
\footnote
{
The model-independent fit to the current Higgs data shows that,
when the bottom- and tau-Yukawa couplings are allowed to vary 
in addition to the gauge-Higgs and top-Yukawa couplings,
the negative top-Yukawa coupling is still allowed at 95\% CL 
due to some collaborative effects from the bottom- and tau-Yukawa
couplings~\cite{higgcision}.
}.

On the other hand, one of the most exciting results from both ATLAS and CMS
in their Run-1 data was the excess in the same-sign dilepton events
with $b$-jets and missing transverse energy
\cite{Aad:2015gdg,Khachatryan:2014qaa,ss2l-ex}. 
The ATLAS collaboration reported a significance of
about $2\sigma$ in the exotic search \cite{Aad:2015gdg}
and the CMS collaboration a significance of about $2.5\sigma$ in
the $t\bar t h$ Higgs search~\cite{Khachatryan:2014qaa}. 
Some people have taken them as the twilight of new physics
beyond the SM (BSM)\cite{ss2l-ph}.

In this work, we focus on the excess observed in
Higgs boson production in association with a top-quark pair ($t\bar t h$). 
In the same sign dilepton channel ($ss2\ell$), 
the best-fit signal strengths are:
$\mu^{\rm ATLAS}_{tth\,,ss2\ell}=2.8\,^{+2.1}_{-1.9}$~\cite{Aad:2015iha} 
and
$\mu^{\rm CMS}_{tth\,,ss2\ell}=5.3\,^{+2.1}_{-1.8}$~\cite{Khachatryan:2014qaa}.
The CMS excess is about $2.5\sigma$ above the SM prediction
while the ATLAS result is still consistent with the SM.
While, the best-fit signal strengths for combined
channels are: 
$\mu^{\rm ATLAS}_{tth}=1.81\pm 0.80$  and
$\mu^{\rm CMS}_{tth}=2.75\pm 0.99 $ 
at $\sqrt{s}$= 7 and 8 TeV \cite{Khachatryan:2014jba}.
Even though the data do not show a significant deviation
from the SM predictions
and there is no obvious indication yet whether
the excess is real,
there are still enough rooms for the 
implication of new physics beyond the SM.
Here we attempt to interpret the mild excess by exploiting the strong
entanglement between the associated Higgs production with a single top quark
($thX$)  and $t\bar t h$ production
in the presence of anomalous top-Yukawa coupling.

%\begin{figure}[th!]
\begin{figure}[t!]
\centering
\includegraphics[width=2.1in]{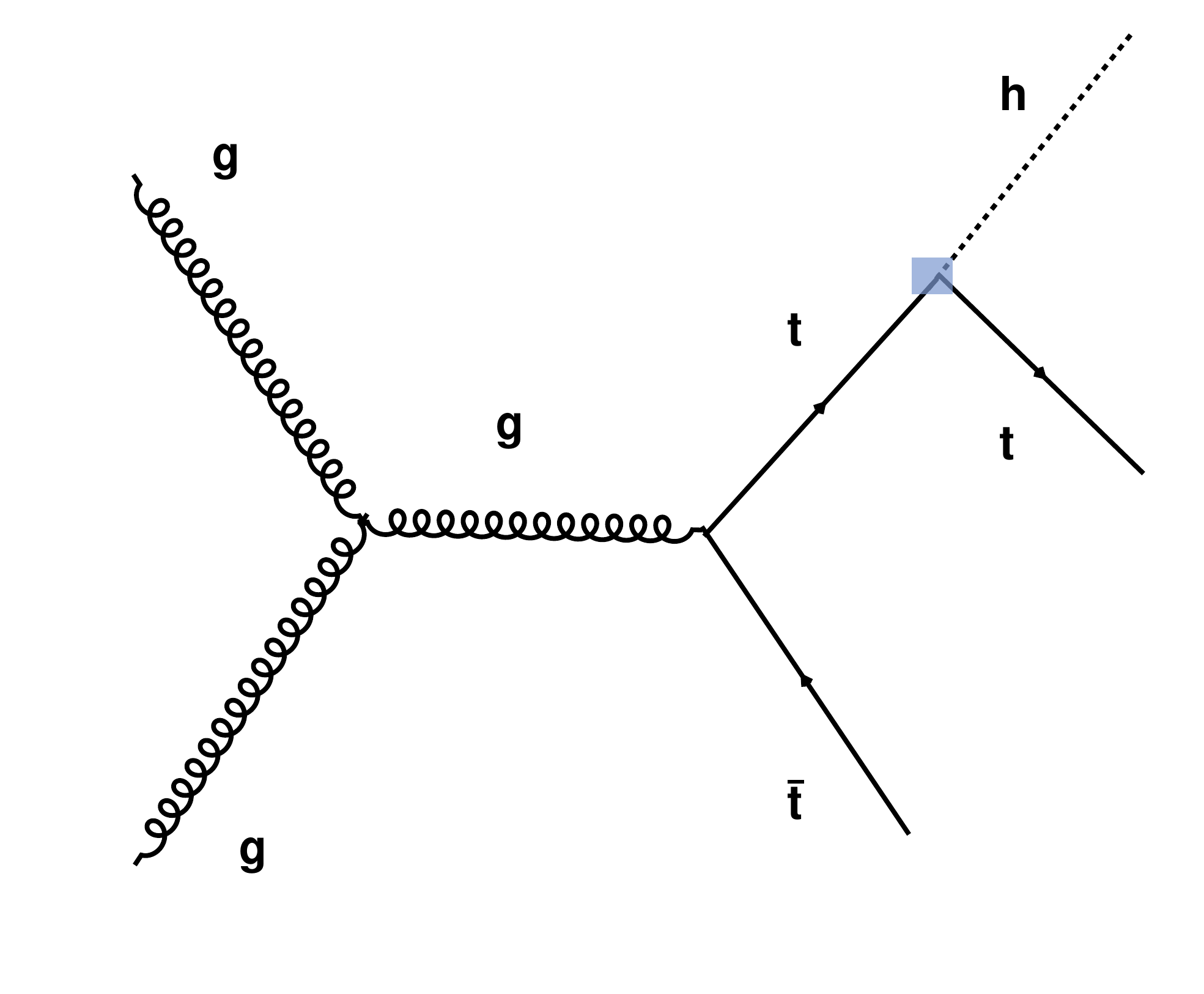}
\includegraphics[width=2.1in]{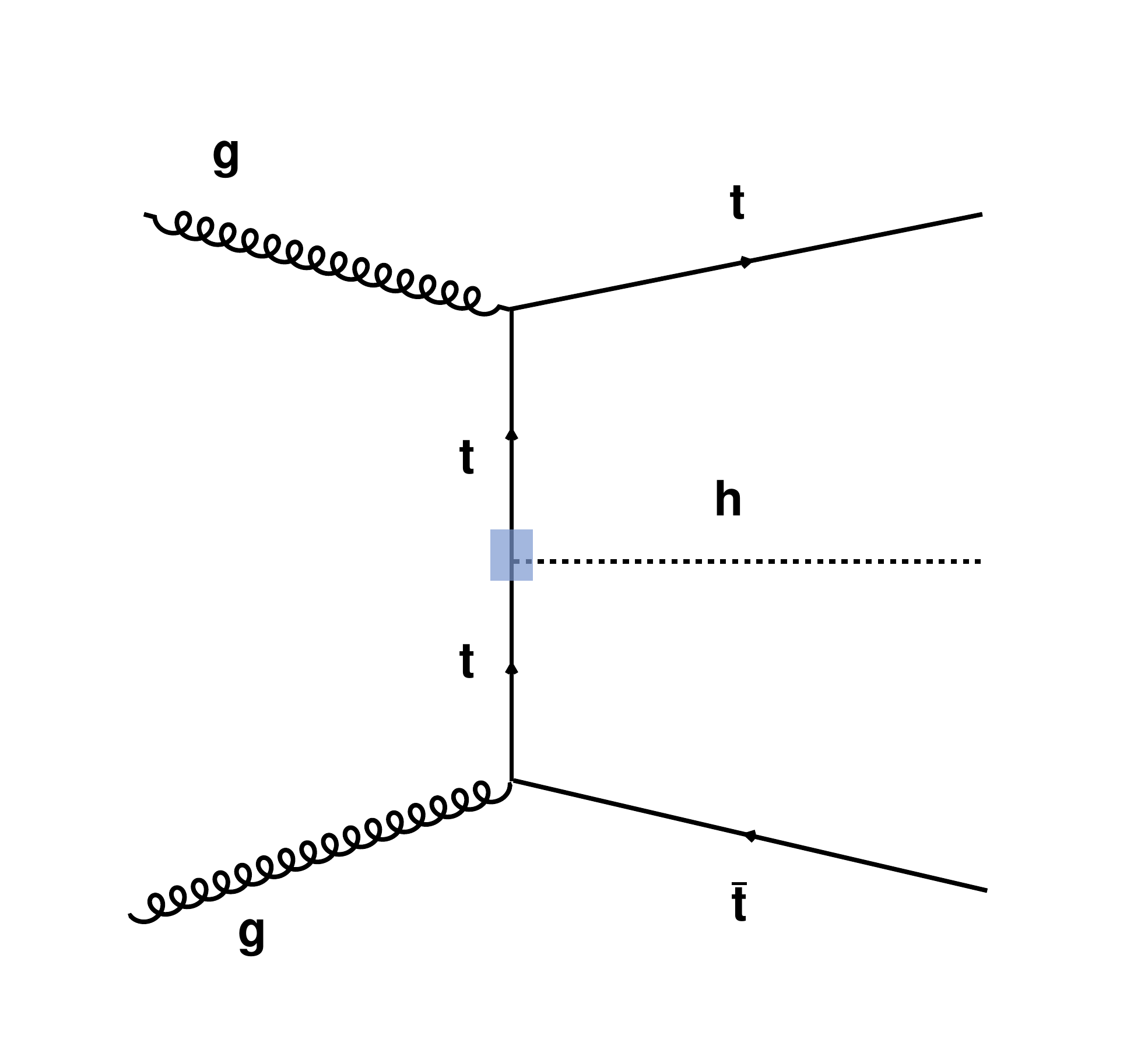}
\includegraphics[width=2.1in]{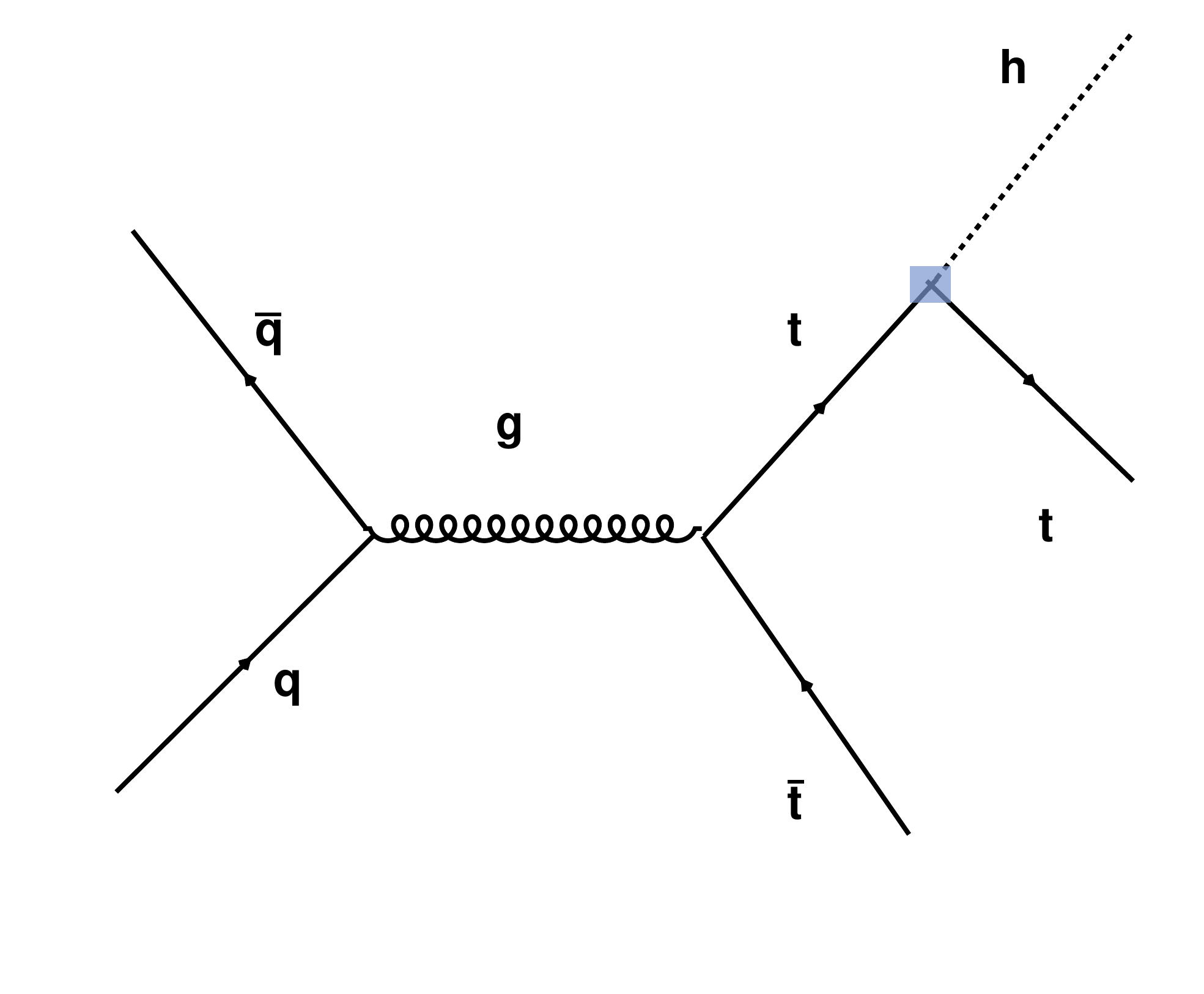}
\caption{\small \label{tth}
Feynman diagrams contributing to $t\bar t h$ production at LO.
}\end{figure}

%\begin{figure}[th!]
\begin{figure}[t!]
\centering
\includegraphics[width=2in, height=1.5in]{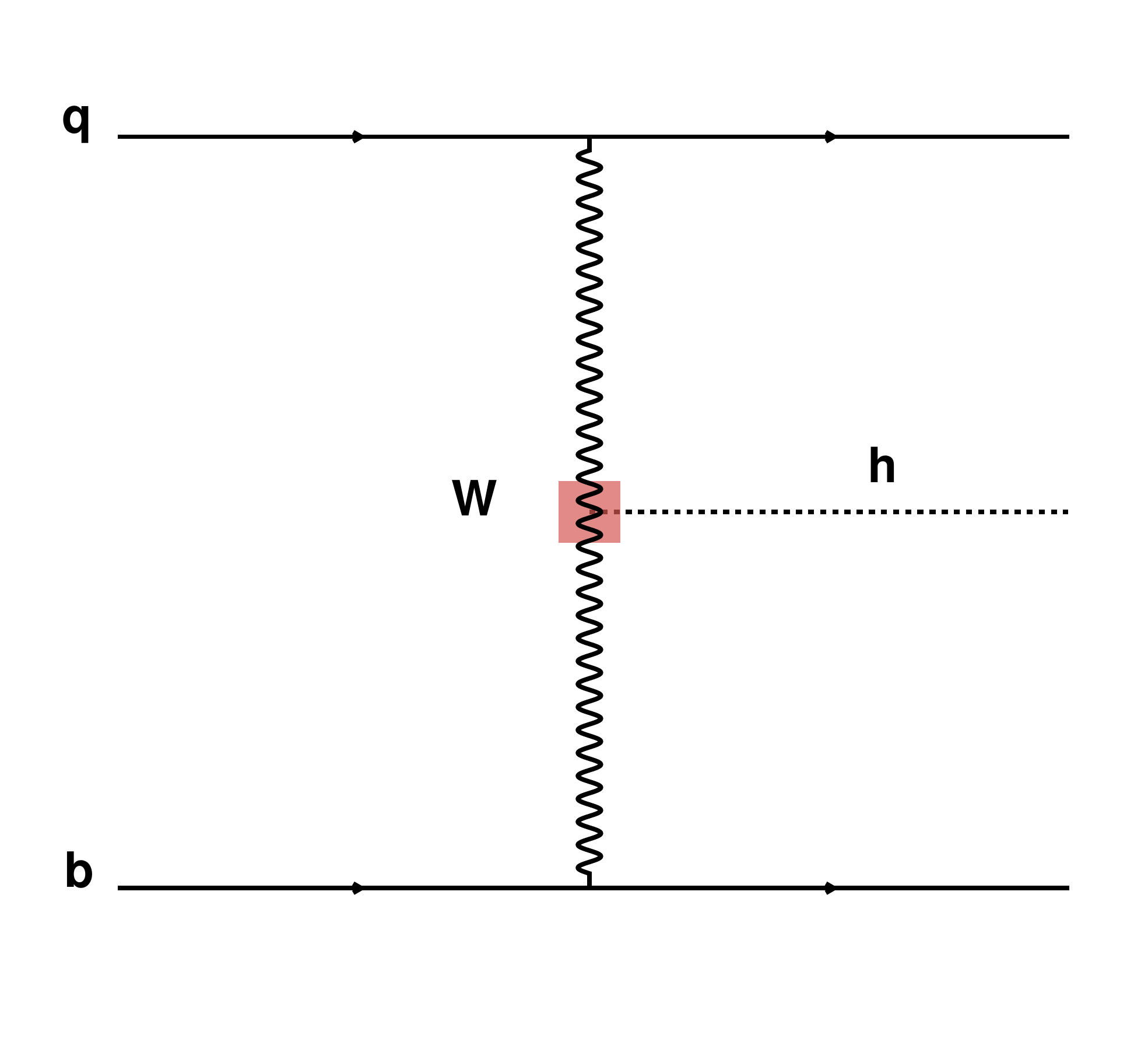}
\includegraphics[width=2in, height=1.5in]{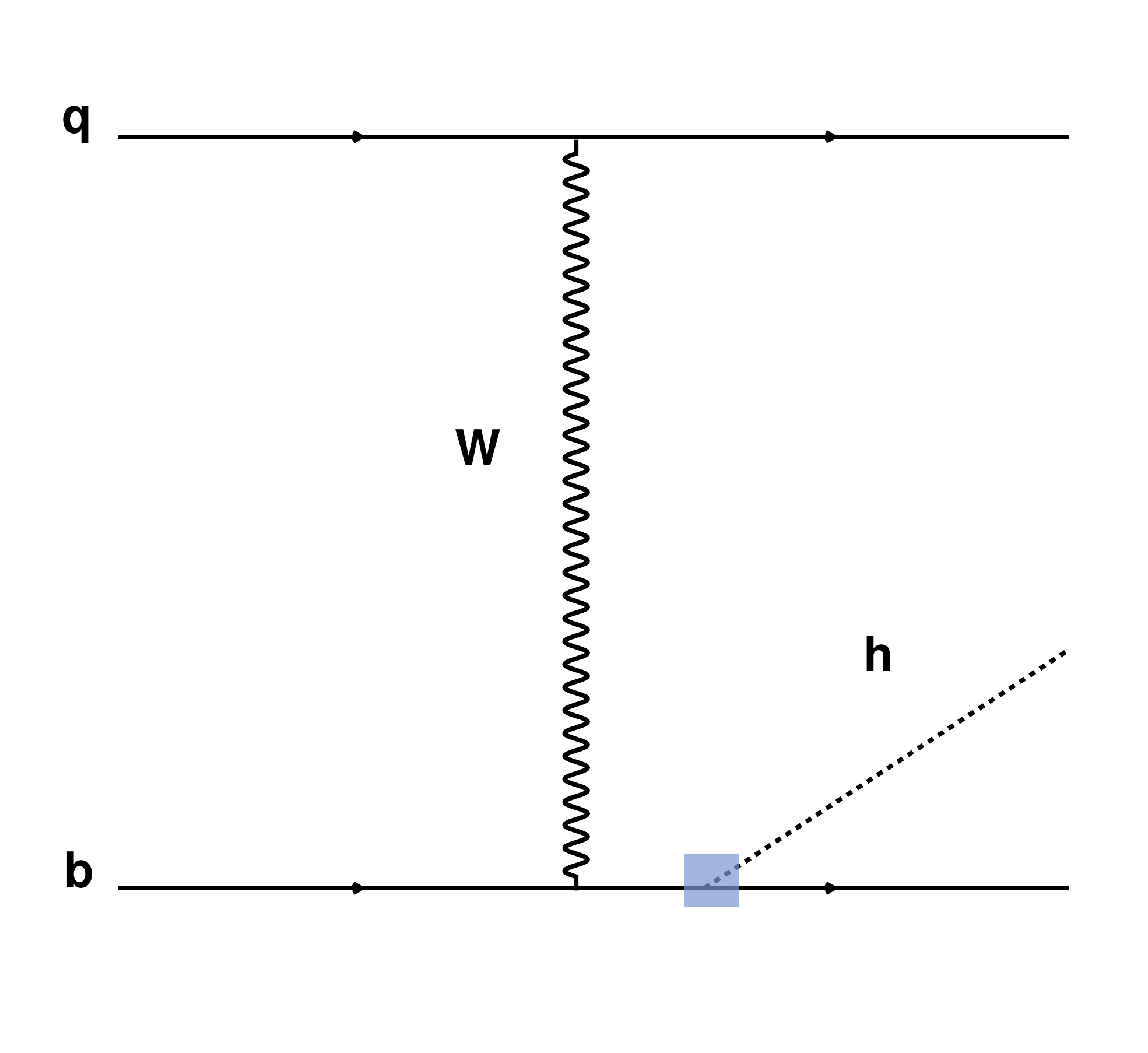}
\caption{\small \label{fig1}
Feynman diagrams contributing to $thX$ production with $X=j$.
}\end{figure}

As well known
$t\bar t h$ production only depends on the absolute value of the
top-Yukawa coupling at the leading order (LO), see Fig.~\ref{tth},
which is similar to gluon-gluon fusion production.
Therefore, the $t\bar{t}h$ cross section is insensitive to the sign
of the top-Yukawa coupling at LO.
Meanwhile, in $thX$ production,
this degeneracy is lifted through the strong
interference between the two main contributions, which are proportional to the
top-Yukawa and the gauge-Higgs couplings, respectively.
Note that we include both $th+X$ and $\bar th +X$
production when we refer to $thX$
with $X$ denoting the accompanied particle(s) produced together 
with $t(\bar t)$ and $h$.
The Feynman diagrams contributing to $thX$ production with $X=j$
($q b \to t h q^\prime$) is depicted in Fig.~\ref{fig1}.
The left diagram 
is proportional to the gauge-Higgs coupling while the right one 
to the top-Yukawa coupling
\footnote{
We neglect the diagram
with the Higgs boson attached to the
bottom-quark leg which is suppressed by the small bottom-Yukawa
coupling.}.  
The interference between the two
diagrams was shown to be significant and induces large
variations in the total cross section with the size and the relative
sign of the Higgs couplings to the gauge bosons and the top quarks. 
It was shown in literature \cite{Chang:2014rfa,singletop.old,th-others} that the
cross section can be enhanced by more than an order of magnitude
when the relative sign of the top-Yukawa coupling 
to the gauge-Higgs coupling is reversed.

In this work,
we perform a detailed study of the influence of $thX$ 
production on $t\bar{t}h $ production
in the presence of the anomalous top-Yukawa coupling.
While assuming the Standard Model (SM) value for the gauge-Higgs coupling,
we vary the top-Yukawa coupling within the allowed range  by the current LHC 
Higgs data.
We consider the Higgs decay modes into multileptons
\footnote{
For earlier proposals to measure the top-Yukawa coupling
through the multilepton modes in $tth$ production,
see Ref.~\cite{tth:multilepton.old}.},
$b\bar b$ and $\gamma\gamma$ putting a particular emphasis 
on  the same sign dilepton events.
We show that
the current ATLAS and CMS analyses of $t\bar t h$
could be significantly
contaminated by the $thX$ processes.
Moreover, the $thX$ processes contribute (or contaminate)
at quite different levels in various detection modes of $tth$,
depending on the value of top-Yukawa coupling, on the cuts
used in each experiment, and on the decay mode of the Higgs
boson.  We shall illustrate such behavior in Sec. III, which is
far more complicated than simply assuming a small constant
level of contamination in all channels.
In addition to explaining the apparent mild excess in $t\bar{t}h$ 
production by entangling $thX$ production,
we also propose how
to disentangle $thX$ production from $t\bar{t}h$  one
at the LHC Run-2. The main objective of this work is to further pin 
down the sign and the size of the top-Yukawa coupling.
To achieve the objective, we point out that
it is crucial to consider the entanglement between $thX$ and $t\bar t h$.

Note that the $\sim 2 \sigma$ excesses were seen in the 
channels of multileptons and $b\bar b$ of ATLAS and in the channels
of multileptons and $\gamma\gamma$ of CMS, but not in the others. 
It may as well be due to statistical fluctuations, but could also be 
due to some specific forms of new physics. Only more data can tell.
%
%In this work, we perform a case study of regarding the mild excess
%being due to some new physics, which manifests in 
%the anomalous top-Yukawa coupling.  
In this work, we perform a case study in which,
through the $thX$ processes,
the contributions of the anomalous top-Yukawa coupling 
to $t\bar t h$ production
manifest non-trivially depending on 
the value of top-Yukawa coupling, on the cuts
used in each experiment, and on the decay mode of the Higgs.
Our case study shows that 
%the current observations
%made in Higgs boson production in association with a top-quark pair
%strongly depends on the Higgs decay modes as well as on the experimet 
%(ATLAS or CMS) and 
the (future) observations related to $t\bar t h$ production
should be carefully made 
without simply assuming a small constant
level of contamination in all channels which is
common to both the ATLAS and CMS experiments.

The organization is as follows. In the next section, we lay down the
formalism and the calculation method. 
In Sec. III, we show the
influence of $thX$ with the anomalous top-Yukawa coupling
on $t\bar{t}h$ for both the ATLAS and CMS Run-1 data.
In Sec. IV, we propose some scenarios to
further disentangle $thX$ from
$t\bar{t}h$ for the LHC Run-2. 
Finally, we discuss and conclude in Sec. V.

\section{Formalism}

\subsection{Processes and Higgs couplings involved}

\begin{figure}[t!]
\centering
\includegraphics[width=2in, height=1.5in]{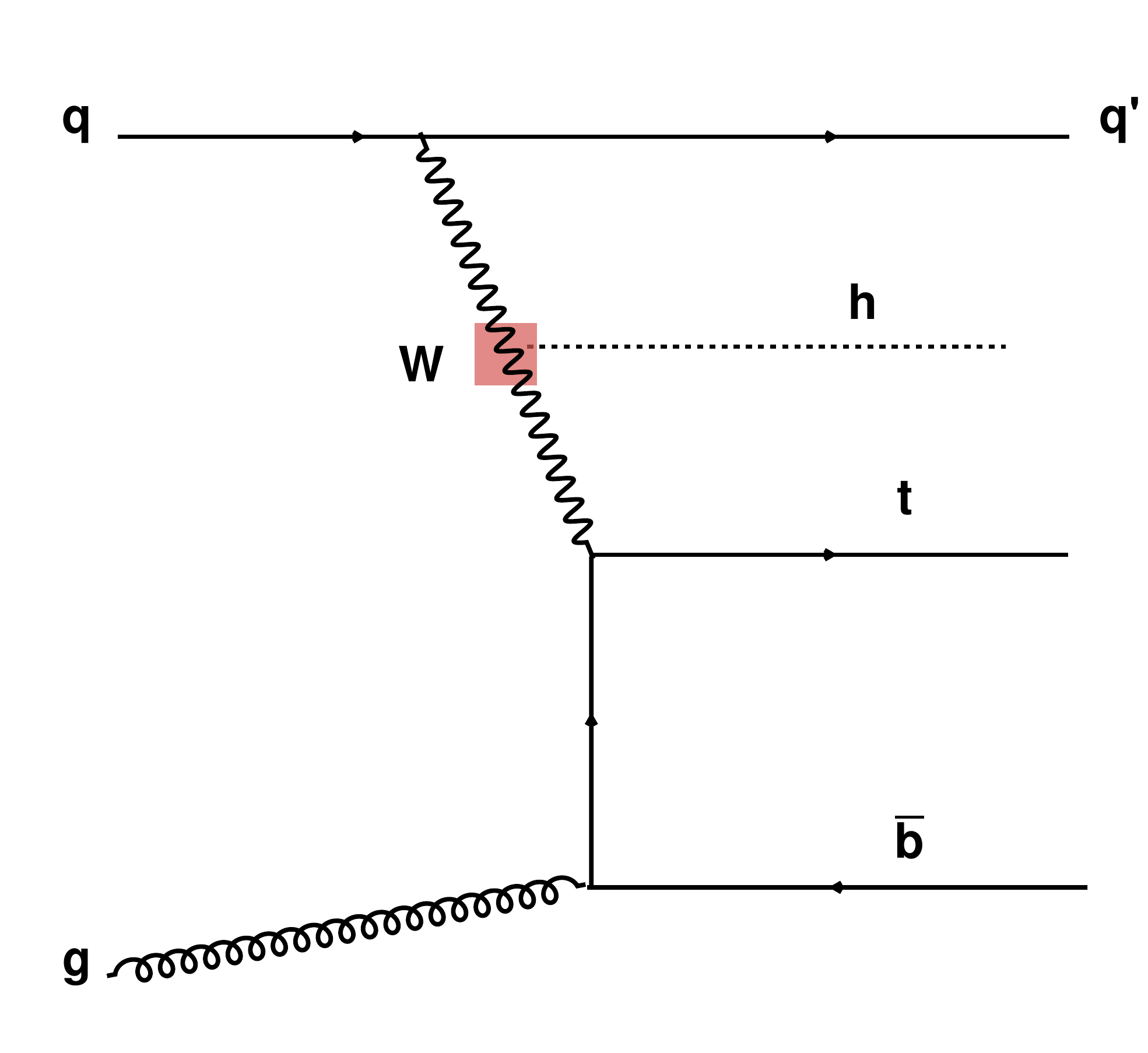}
\includegraphics[width=2in, height=1.5in]{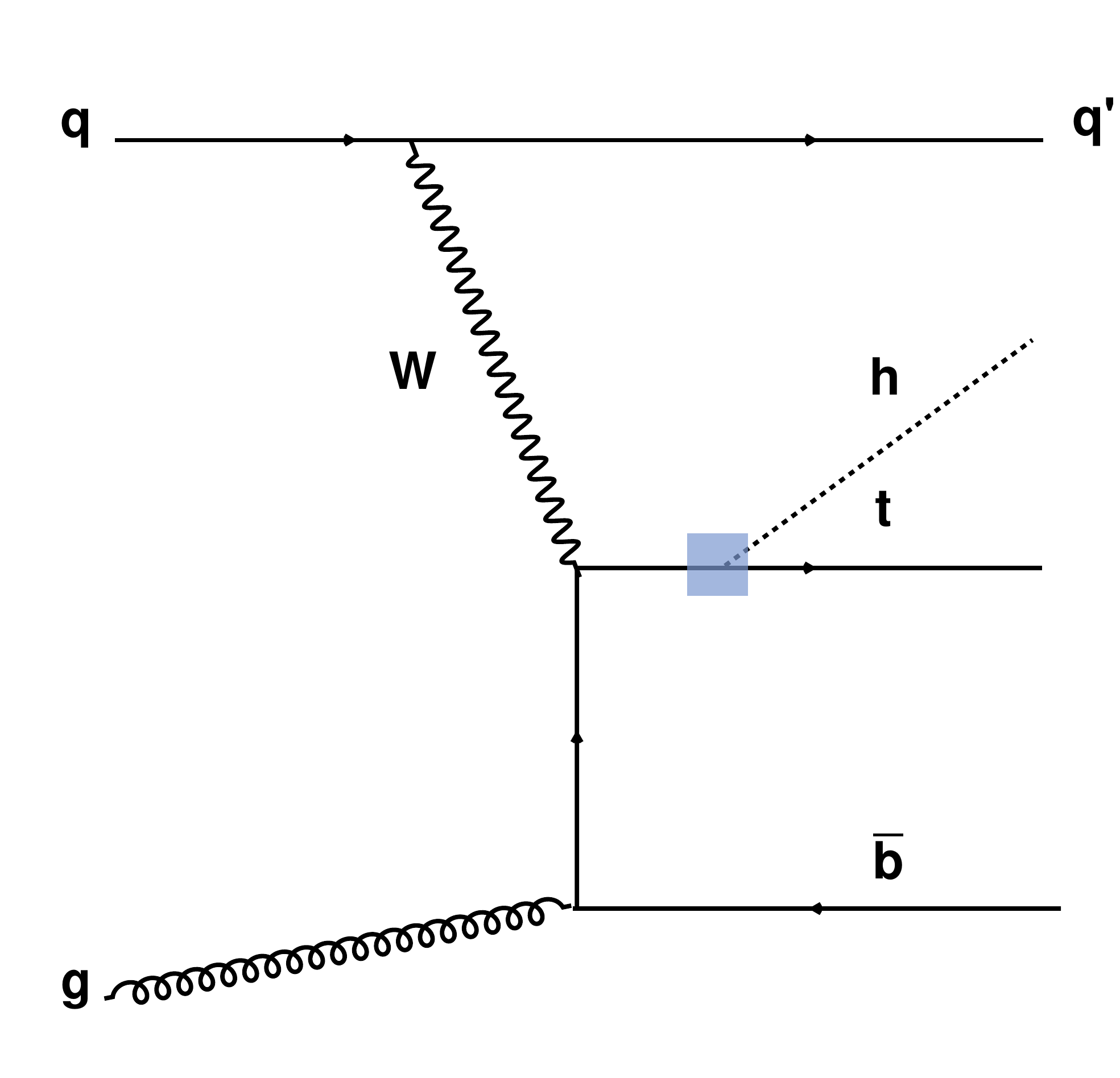}
\caption{\small \label{fig2}
Feynman diagrams contributing to $thX$ production with $X=jb$.
}\end{figure}

\begin{figure}[t!]
\centering
\includegraphics[width=2.1in]{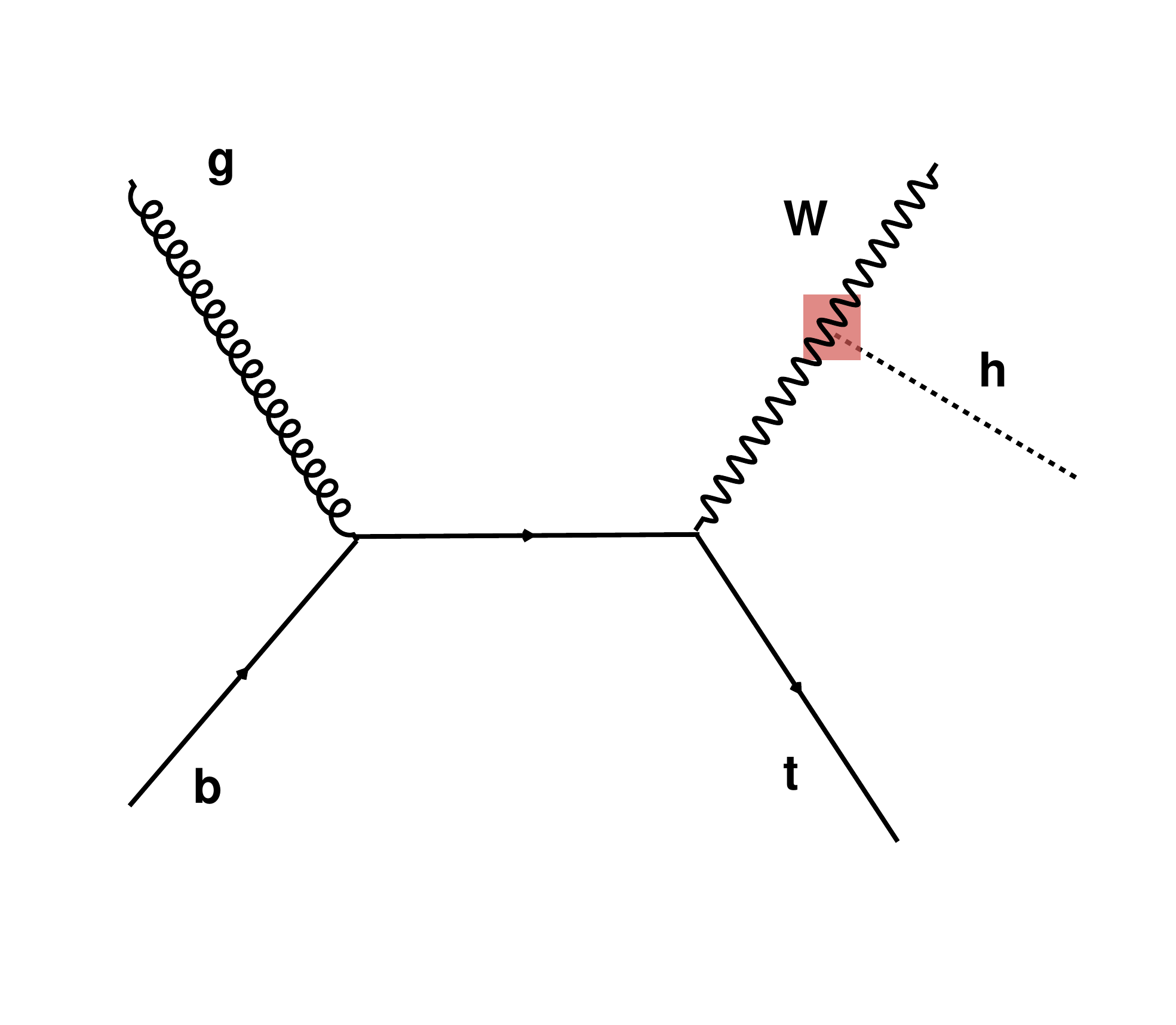}
\includegraphics[width=2.1in]{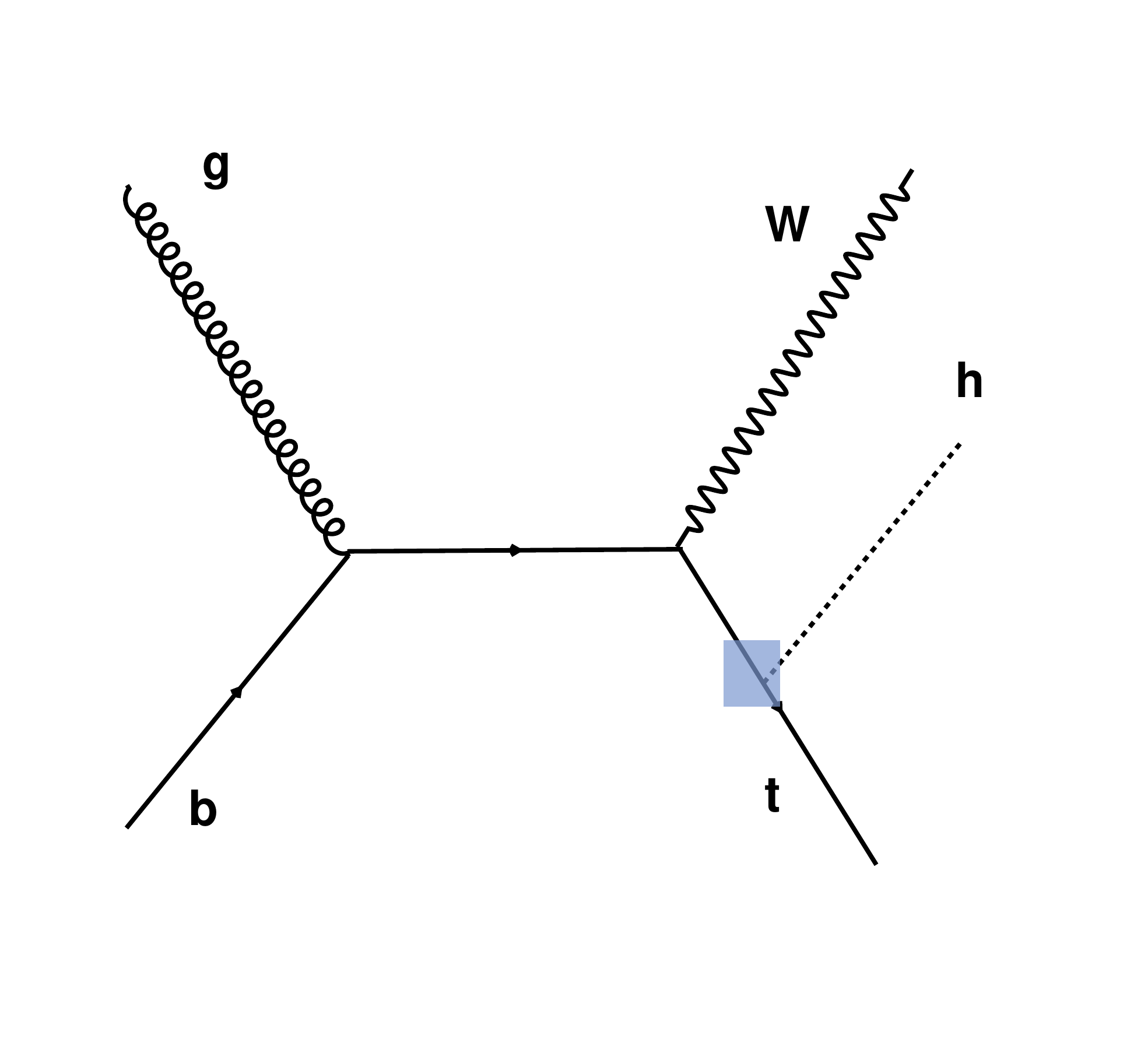}
\caption{\small \label{fig3}
Feynman diagrams contributing to $thX$ production with $X=W$.
}\end{figure}

We consider two types of production processes for the Higgs boson and the top
quark. The first one is the associated production of the Higgs with
a pair of top quarks, see Fig.~\ref{tth}. The second one is
the associated Higgs production with a single top quark plus anything 
else: $thX$ production with 
$X=j\,(qb\to thq'),\ jb\,(qg\to thq'b),\ W\,(gb\to thW)$
\footnote{
In this work,
we ignore the $s$-channel $thX$ process with $X=b$ 
$(q\bar q' \to thb)$
because its production cross section is much smaller compared to
other processes with $X=j\,,jb\,,W$.
}, 
see Figs.~\ref{fig1} -- \ref{fig3}
in which we have marked the vertices of $hWW$ and 
$ht\bar{t}$ with squares.
In $t\bar{t}h$ production,
the production cross section only depends on the square of the 
top-Yukawa coupling. However, in $thX$ production,
the cross sections depend on the size of the gauge-Higgs and 
top-Yukawa couplings and the relative sign between them.

%\begin{figure}[th!]

%\begin{figure}[th!]

%\begin{figure}[th!]
%\centering
%\includegraphics[width=2.1in]{figure/thb_f1.pdf}
%\includegraphics[width=2.1in]{figure/thb_f2.pdf}
%\includegraphics[width=2.1in]{figure/thb_f3.pdf}
%\caption{\small \label{fig4}
%Feynman diagrams contributing $thX$ production with $X=b$.
%}
%\end{figure}

In fact, the process $qg\to thjb$ is a part of the NLO 
QCD corrections to $qb \to thj$ when
the momentum of the final $b$ quark in $thjb$ is integrated out.
In our work, using MadGraph5@NLO, we calculate the cross section for the 
$qg\to thjb$ process at NLO
adopting the four-flavor scheme.
And then, we define $thj$ and $thjb$ productions by introducing a set of
separation cuts:
$p_T^j>10$ GeV, $|\eta_j|<5$, $p_T^b>30$ GeV, $|\eta_b|<2.5$.
Naturally, the low (high) $p_T^b$ region is taken for
$thj$ ($thjb$) production.
We obtain $\sigma(thj) = 11.2\,(43.0) $ fb and $\sigma(thjb) = 5.77\,(23.6)$ fb
at the LHC with $\sqrt{s}= 8\,(13)$ TeV. We note that the sum
$\sigma(thj)+\sigma(thjb)= 17.0\,(66.6)$ fb
agrees well with the NLO cross sections found in the literature.
As will be shown, the contributions of $thj$ and $thjb$ to the
accumulated signal strengths strongly depend on the Higgs decay channels and 
experiment cuts chosen.
In this way, we properly reflect the
different kinematic signatures of $thj$ and $thjb$
which could be lost if we do not introduce the separation.
On the other hand, since the NLO QCD corrections 
for both $ pp\to t\overline{t}h $ and $ pp\to thW $ are relatively 
large compared with $ pp\to thj $, we multiply the corresponding 
$K$ factors to the LO cross sections for each of them.

Without loss of generality,
one can write the gauge-Higgs and Yukawa couplings 
of the Higgs boson $h$ as
\footnote{
In this work, we assume that the Higgs boson $h$ 
is a generic CP-even state which arbitrarily couples to the SM 
and BSM particles.}

\begin{eqnarray}
{\cal L}_{hVV} &=& g m_W \left( g_{hWW} W^+_\mu W^{-\mu} + 
 g_{hZZ} \frac{1}{2 c_W^2} Z_\mu Z^\mu \right ) h \,,  \label{hvv}\\
{\cal L}_{hff} &=& - \sum_{f = t,b,c,\tau} \, \frac{g m_f}{2 m_W} \,
g^S_{hff} \, \bar f \, f \,h \;. 
\label{hff}
\end{eqnarray}
Here only the gauge-Higgs coupling $g_{hWW}$ and the top-Yukawa couplings
are relevant to the $t\bar t h$ and $thX$ production processes
shown in Figs.~\ref{tth}--\ref{fig3}. 
We note $g_{hWW} = g_{hZZ} = g^S_{hff} = 1$ in the SM. 

In order to calculate the event rates we have to consider the decay
branching ratios of the Higgs boson, which depend on 
$g_{hWW}$, $g_{hZZ}$, $g^{S}_{htt,hbb}$
and a few more couplings, including $h\tau\tau$, $hcc$, $h\gamma\gamma$,
and $hgg$. 
The amplitude for the decay process
$h \rightarrow \gamma\gamma$ can be written as
\begin{equation} \label{hipp}
{\cal M}_{h\gamma\gamma}=-\frac{\alpha m_{h}^2}{4\pi\,v}
\,S^\gamma(m_{h})\,
\left(\epsilon^*_{1\perp}\cdot\epsilon^*_{2\perp}\right)\,,
\end{equation}
where $k_{1,2}$ are the momenta of the two photons and
$\epsilon_{1,2}$ the wave vectors of the corresponding photons
with
$\epsilon^\mu_{1\perp} = \epsilon^\mu_1 - 2k^\mu_1 (k_2 \cdot
\epsilon_1) / m^2_{h}$ and $\epsilon^\mu_{2\perp} = \epsilon^\mu_2 -
2k^\mu_2 (k_1 \cdot \epsilon_2) / m^2_{h}$.
Retaining only the dominant loop
contributions from the third--generation fermions and $W^\pm$,
and 
including some additional loop contributions from new particles,
the scalar form factor is given by
\begin{eqnarray}
\label{eq:hrr}
S^\gamma(m_{h})&=&2\sum_{f=b,t,\tau} N_C\,
Q_f^2\, g^{S}_{hff}\,F_{sf}(\tau_{f}) 
- g_{_{hWW}}F_1(\tau_{W}) 
+ \Delta S^\gamma \,, 
\end{eqnarray}
where $\tau_{x}=m_{h}^2/4m_x^2$, $N_C=3$ for quarks and $N_C=1$ for
tau leptons, respectively.
For the loop functions of $F_{sf,1}(\tau)$, 
we refer to, for example, Ref.~\cite{Lee:2003nta}.
The additional contributions $\Delta S^\gamma$ 
are due to additional particles running in the loop. In the SM,
$g^S_{hff}= g_{hWW}=1$ and $\Delta S^\gamma=0$.
%%%
%%%
Similarly, the amplitude for the decay process
$h \rightarrow gg$ can be written as
\begin{equation} \label{higg}
{\cal M}_{Hgg}=-\frac{\alpha_s\,m_{h}^2\,\delta^{ab}}{4\pi\,v}
\, S^g(m_{h}) \,
\left(\epsilon^*_{1\perp}\cdot\epsilon^*_{2\perp}\right) \,,
\end{equation}
where $a$ and $b$ ($a,b=1$ to 8) are indices of the eight $SU(3)$
generators in the adjoint representation.
Including some additional loop contributions from new particles,
the scalar form factor is given by
\begin{eqnarray}
S^g(m_{h})&=&\sum_{f=b,t}
g^{S}_{hff}\,F_{sf}(\tau_{f}) +  
\Delta S^g\,.
\end{eqnarray}
In the SM, $g^S_{hff}=1$ and $\Delta S^g=0$.
In the decays of the Higgs boson, we can see that the partial width
into $b\bar b$ depends on $g_{hbb}$, that into $WW^*$ and $ZZ^*$ depends
on $g_{hWW,hZZ}$, and that into $\gamma\gamma$ and $gg$ depends implicitly
on all $g_{hWW}$, $g_{htt}^{S}$, $g_{hbb}^{S}$, and $g_{h\tau\tau}^{S}$.

The dependence of the production cross sections and the decay branching
ratios on $g_{hWW}$ and $g^{S}_{hff}$ has been explicitly shown in the above
equations. 
Since we are primarily interested in 
size of the gauge-Higgs and top-Yukawa couplings and
the relative sign between them,
for bookkeeping purpose, we use the following simplified notations
\begin{equation}
\label{eq:notation}
C_v  \equiv g_{hWW}=g_{hZZ}\,, \qquad
C^{S}_t  \equiv g^{S}_{htt}\,, \qquad
C^{S}_b  \equiv g^{S}_{hbb}\,.
\end{equation}

We shall show the anomalous top-Yukawa coupling effects on 
$t\bar{t}h$ and $thX$ production at the LHC in the next section.

\subsection{Signal strengths}
First we note that signal strengths depend on the 
decay modes of the top quark and the Higgs boson, as well as their 
production mechanisms. For a choice of experimentally-defined
decay mode ${\cal D}$, and taking into account the $thX$ production processes,
we define the signal strength $\mu(t\bar t h)$ 
with respect to the SM $t\bar{t}h$ production as follows
\begin{equation}
\label{mu-tth}
\mu(t\bar{t}h)=
\frac{
\eta_1\sigma(t\bar{t}h)B(t\bar{t}h\to{\cal D})+
\sum_{X=j,jb,W}\eta_X\sigma(thX)B(thX\to{\cal D})}
{\eta_1^{\rm SM}\sigma(t\bar{t}h)_{\rm SM}B(t\bar{t}h\to{\cal D})_{\rm SM}}\,,
\end{equation}
where $\sigma(t\bar{t}h)=\sigma(pp\to t\bar{t}h)$ and
$\sigma(thX)=\sigma(pp\to thX)+\sigma(pp\to \bar{t}hX)$ are  understood.
The detection efficiencies $\eta$'s depend
on the experimental apparatuses and cuts
for the specific production and decay mode.
By introducing the cross-section ratios
\begin{eqnarray}
&&
R(t\bar{t}h)\equiv\frac{\sigma(t\bar{t}h)}{\sigma(t\bar{t}h)_{\rm SM}}\,, \ \ \
R(thj)\equiv\frac{\sigma(thj)}{\sigma(t\bar{t}h)_{\rm SM}}\,, \ \ \
\nonumber \\[2mm] &&
R(thjb)\equiv\frac{\sigma(thjb)}{\sigma(t\bar{t}h)_{\rm SM}}\,, \ \ \
R(thW)\equiv\frac{\sigma(thW)}{\sigma(t\bar{t}h)_{\rm SM}}\,,
\label{R-thX}
\end{eqnarray}
and the ${\cal D}$-dependent detection-efficiency ratios
\begin{eqnarray}
&&
\epsilon_1\equiv\frac{\eta_1B(t\bar{t}h\to{\cal D})}
{\eta_1^{\rm SM}B(t\bar{t}h\to{\cal D})_{\rm SM}}\,, \ \ \
\epsilon_2\equiv\frac{\eta_jB(thj\to{\cal D})}
{\eta_1^{\rm SM}B(t\bar{t}h\to{\cal D})_{\rm SM}}\,, \ \ \
\nonumber \\[2mm] &&
\epsilon_3\equiv\frac{\eta_{jb}B(thjb\to{\cal D})}
{\eta_1^{\rm SM}B(t\bar{t}h\to{\cal D})_{\rm SM}}\,, \ \ \
\epsilon_4\equiv\frac{\eta_{W}B(thW\to{\cal D})}
{\eta_1^{\rm SM}B(t\bar{t}h\to{\cal D})_{\rm SM}}\,, 
\label{ei}
\end{eqnarray}
one may have
\begin{equation}
\mu(t\bar{t}h)=
\epsilon_1\,R(t\bar{t}h)+
\epsilon_2\,R(thj)+
\epsilon_3\,R(thjb)+
\epsilon_4\,R(thW)\,.
\end{equation}
We note that $\epsilon_1=R(t\bar{t}h)=1$ 
in the SM limit of $C_v=1$ and $C_t^S=+1$
and $\mu(t\bar t h)$ is always larger than $1$ 
due to the entanglement of $thX$ production.
Our main task is to calculate 
the cross section ratios $R$'s in 
the presence of anomalous top-Yukawa coupling
and
the detection-efficiency ratios $\epsilon_{1,2,3,4}$ for various
top-quark and Higgs-boson decay modes.

\begin{table}[t!]
%\begin{table}[th!]
\caption{\small \label{CMS-ATLAS1-table}
The best-fit values for the category-dependent
signal strengths $\mu_{tth}^{\rm CMS}$ and
$\mu_{tth}^{\rm ATLAS}$ coming from
the CMS \cite{Khachatryan:2014qaa} and
  ATLAS \cite{Aad:2015iha}\cite{Aad:2014lma}\cite{Aad:2015gra} searches,
respectively,
  for the associated production of the Higgs boson with a top quark
  pair at $\sqrt{s}$= 7 and 8 TeV for $ m_{h}=125.6 $ GeV (CMS) / 125 GeV
(ATLAS). }
\medskip
\begin{ruledtabular}
\begin{tabular}{lcc}
& \multicolumn{1}{c}{CMS $ t\bar{t}h $ channel}& \multicolumn{1}{c}{ATLAS $
t\bar{t}h $ channel}  \\
%Category & Best-fit $\mu_{tth}^{\rm CMS}$  
%         & Best-fit $\mu_{tth}^{\rm ATLAS}$ \\
Category & $\mu_{tth}^{\rm CMS}$  
         & $\mu_{tth}^{\rm ATLAS}$ \\
\hline
$ \gamma\gamma $ &  $ +2.7\,^{+2.6}_{-1.8} $ &  $ +1.3\,^{+3.3}_{-2.1} $\\
$ b\bar{b} $ &  $ +0.7\,^{+1.9}_{-1.9} $ &  $ +1.5\,^{+1.1}_{-1.1} $  \\
$ \tau_{h}\tau_{h} $ &  $ -1.3\,^{+6.3}_{-5.5} $&-- \\
$ 2\ell 1\tau_{h} $ & -- &$ -0.9\,^{+3.1}_{-2.0} $ \\
$ 1\ell 2\tau_{h} $ & -- &$ -9.6\,^{+9.6}_{-9.7} $ \\
$ 4\ell $ &  $ -4.7\,^{+5.0}_{-1.3} $&  $ +1.8\,^{+6.9}_{-2.0} $ \\
$ 3\ell $ &  $ +3.1\,^{+2.4}_{-2.0} $&  $ +2.8\,^{+2.2}_{-1.8} $ \\
$ ss2\ell $ &  $ +5.3\,^{+2.1}_{-1.8} $ &  $ +2.8\,^{+2.1}_{-1.9} $ \\
\end{tabular}
\end{ruledtabular}
\end{table}

\section{$thX$ production with the anomalous top-Yukawa coupling}
Both the CMS \cite{Khachatryan:2014qaa} and ATLAS
\cite{Aad:2014lma,Aad:2015gra,Aad:2015iha} collaborations
have published the results of
their searches for the associated production of the Higgs boson with a
top-quark pair via different Higgs decay channels at $\sqrt{s}$= 7 and
8 TeV. We summarize their best-fit results in
Table~\ref{CMS-ATLAS1-table}.  
Since the experimental
uncertainties in the hadronically-decaying $\tau$ 
and $4\ell$ categories are too large at this stage,
we shall focus only on the $ss2\ell$, $ 3\ell$, 
$\gamma\gamma$ and $b\bar{b} $ categories in our analysis below. 
In the $\gamma\gamma$ category for $h\to\gamma\gamma$,
both CMS
\cite{Khachatryan:2014qaa} and ATLAS \cite{Aad:2014lma} included all
the decay modes of a top-quark pair: semileptonic 
($t\bar{t}\rightarrow l\nu jjbb $), leptonic ($ t\bar{t}\rightarrow
l\nu\l\nu bb $), and hadronic ($ t\bar{t}\rightarrow jjjjbb $) modes. 
On the other hand, 
in the $b\bar b$ category for $h\to b \bar b$,
both CMS
\cite{Khachatryan:2014qaa} and ATLAS \cite{Aad:2015gra} 
considered only the
semileptonic and leptonic decay modes of the top-quark pair.
Finally, in the categories of $ss2\ell $ and $ 3\ell $ for
$h\rightarrow~{\rm multileptons}$, 
both CMS \cite{Khachatryan:2014qaa} and ATLAS \cite{Aad:2015iha}
included only the semileptonic decay mode of the top-quark pair.

In order to perform a detailed study of the influence  of 
$thX$ production with anomalous top-Yukawa coupling on 
$ t\bar{t}h $ production,
we simulate both the
$ thX$ and $ t\bar{t}h $ processes and
generate events by MadGraph5 \cite{Alwall:2014hca}, perform parton
showering and hadronization by Pythia 8.1 \cite{Sjostrand:2007gs}, and
employ the detector simulations by Delphes 3 \cite{delphes3}. We use
NN23LO1 for parton distribution functions with different
renormalization/factorization scales which we shall show below. We
follow the selection cuts and detector efficiencies of the CMS
\cite{Khachatryan:2014qaa} and ATLAS
\cite{Aad:2014lma,Aad:2015gra,Aad:2015iha} $t\bar th$ searches. We
summarize the signatures of the search channels used in the $t\bar th$
analysis for CMS in Table~\ref{CMS2-table} and for ATLAS in
Table~\ref{ATLAS2-table}.

%\begin{table}[bth!]
\begin{table}[t!]
\caption{\label{CMS2-table} \small
The signature of the search channels used in the $tth$  analysis for 
CMS.
}
\medskip
\begin{ruledtabular}
\begin{tabular}{|l|c|c|c|}
Category & $ t\overline{t}h $ decay modes & Signature & Background \\
\hline \hline
$ h\rightarrow b\overline{b} $ & Semileptonic & 1 $ e/ \mu $, $P_{T} > $ 30 GeV &$t\bar{t} +$ jets\\
 & ($ t\overline{t}h\rightarrow l\nu jjbbbb $) & $\geq$ 4 jets + $\geq$ 2b-tags, $P_{T} > $ 30 GeV &$t\bar{t} +W/Z$ \\
\cline{2-3}
 & Leptonic & 1 $ e/ \mu $, $P_{T} > $ 20 GeV &Single $t$\\
 & ($ t\overline{t}h\rightarrow l\nu l\nu bbbb $) & 1 $ e/ \mu $, $P_{T} > $ 10 GeV &$W/Z+$jets\\
 & & $\geq$ 3 jets + $\geq$ 2b-tags, $P_{T} > $ 30 GeV&Diboson \\
\hline
$ h\rightarrow\gamma\gamma $ & Semileptonic & 2$\gamma$, $ P_{T} > m_{\gamma\gamma}/2 $ (25) GeV for $ 1^{st} $($ 2^{nd} $)&$t\bar{t} +$ jets \\ 
 & ($ t\overline{t}h\rightarrow l\nu jjbb\gamma\gamma $) & { $\geq$ }1 $ e/ \mu $, $P_{T} > $ 20 GeV &$t\bar{t} +W/Z$\\
 & Leptonic & $\geq$ 2 jets + $\geq$ 1b-tags, $P_{T} > $ 25 GeV &Single $t$\\
 & ($ t\overline{t}h\rightarrow l\nu l\nu bb\gamma\gamma $) & & \\
\cline{2-3} 
 & Hadronic &  2$\gamma$, $ P_{T} > m_{\gamma\gamma}/2 $ (25) GeV for $ 1^{st} $($ 2^{nd} $) & \\
 & ($ t\overline{t}h\rightarrow jjjjbb\gamma\gamma $) & 0 $ e/ \mu $, $P_{T} > $ 20 GeV & \\ 
 & &  $\geq$ 4 jets + $\geq$ 1b-tags, $P_{T} > $ 25 GeV & \\
\hline
\textbf{ $h \rightarrow$ Leptons} & Same-Sign Dilepton & 2 $ e/ \mu $, $P_{T} > $ 20 GeV &$t\bar{t} W$ \\
$ h\rightarrow WW $ & ($ t\overline{t}h\rightarrow l^{\pm}\nu l^{\pm}[\nu]jjj[j]bb $) & $\geq$ 4 jets + $\geq$ 1b-tags, $P_{T} > $ 25 GeV &$t\bar{t} Z/\gamma^*$  \\
\cline{2-3}
$ h\rightarrow\tau\tau $ & 3 Leptons & 1 $ e/ \mu $, $P_{T} > $ 20 GeV  &$t\bar{t} WW$\\
$ h\rightarrow ZZ $ & ($ t\overline{t}h\rightarrow l\nu l[\nu]l[\nu]j[j]bb $) & 1 $ e/ \mu $, $ P_{T} > $ 10 GeV &$t\bar{t} \gamma$ \\ 
 & & 1 $ e(\mu) $, $ P_{T} > $ 7(5) GeV &$WZ$\\
 & & $\geq$ 2 jets + $\geq$ 1b-tags, $P_{T} > $ 25 GeV &$ZZ$\\
\end{tabular}
\end{ruledtabular}
\end{table}

%\begin{table}[bth!]
\begin{table}[t!]
\caption{\label{ATLAS2-table} \small
The signature of the search channels used in the $tth$  analysis for ATLAS.
}
\medskip
\begin{ruledtabular}
\begin{tabular}{|l|c|c|c|}
Category & $ t\overline{t}h $ decay modes & Signature &Background \\
\hline \hline
$ h\rightarrow b\overline{b} $ & Semileptonic & 1 $ e/ \mu $, $P_{T} > $ 25 GeV, $ \bigtriangleup R < $ 0.15&$t\bar{t}+$jets \\
 & ($ t\overline{t}h\rightarrow l\nu jjbbbb $) & $\geq$ 4 jets + $\geq$ 2b-tags 
&$t\bar{t}+V$\\
\cline{2-3} 
 & Leptonic & 1 $ e/ \mu $, $P_{T} > $ 25 GeV & $V+$jets\\
 & ($ t\overline{t}h\rightarrow l\nu l\nu bbbb $) & 1 $ e/ \mu $, 
$P_{T} > $ 15 GeV & ($V=W,Z$)\\
 & & $\geq$ 2b-tags & \\
\hline
$ h\rightarrow\gamma\gamma $ & Semileptonic & {2$\gamma$, $ E_{T} > 0.35(0.25)\times m_{\gamma\gamma} $ for $ 1^{st} $($ 2^{nd} $) } & \\
 & ($ t\overline{t}h\rightarrow l\nu jjbb\gamma\gamma $) & $\geq$ 1 $ e/ \mu $, { $ E_{T}(e) > $ 15 GeV, $P_{T}(\mu) > $ 10 GeV } & \\ 
 & Leptonic & $\geq$ 1 b-tags & \\
 & ($ t\overline{t}h\rightarrow l\nu l\nu bb\gamma\gamma $) & & \\
\cline{2-3} 
 & Hadronic & { 2$\gamma$, $ E_{T} > 0.35(0.25)\times m_{\gamma\gamma} $ for $ 1^{st} $($ 2^{nd} $) } & \\ 
 & ($ t\overline{t}h\rightarrow jjjjbb\gamma\gamma $) & { 0 $ e/ \mu $ } & \\
 & & { $\geq$ 5 jets + $\geq$ 1 b-tags, $P_{T} > $ 25 GeV } & \\
\hline
\textbf{ $h\rightarrow$ Leptons} & Same-Sign Dilepton & 
(sub)leading lepton : 2 $ e/ \mu $, $P_{T} > $ 25(20) GeV&$t\bar{t}+$jets \\
$ h\rightarrow WW $ & ($ t\overline{t}h\rightarrow l^{\pm}\nu l^{\pm}[\nu]jjj[j]bb $) & $\geq$ 4 jets + $\geq$ 1b-tags, { $P_{T} > $ 25 GeV } &$t\bar{t}+V$\\
\cline{2-3}
$ h\rightarrow\tau\tau $ & 3 Leptons & {1 $ e/ \mu $, $P_{T} > $ 25 GeV } & Diboson\\
$ h\rightarrow ZZ $ & ($ t\overline{t}h\rightarrow l\nu l[\nu]l[\nu]j[j]bb $) & {1 $ e/ \mu $, $P_{T} > $ 20 GeV } & \\ 
 & & { 1 $ e/ \mu $, $P_{T} > $ 10 GeV } & \\
 & & $\geq$ 4 jets + $\geq$ 1 b-tags, {$P_{T} > $ 25 GeV } & \\ 
 & & (or 3 jets + $\geq$ 2 b-tags, { $P_{T} > $ 25 GeV } ) & \\
\end{tabular}
\end{ruledtabular}
\end{table}

We calculate the $ t\bar{t}h $ production cross section with the factorization
($\mu_{F}$) and renormalization ($\mu_{R}$) scales set at $m_{t}+m_{h}/2 $ 
in the four-flavor scheme. 
On the other hand, in computing the production cross sections for
$thX$, we include the $t$-channel $thj$ and $thjb$ processes and 
the $thW$ process, but ignore the $s$-channel $thb$ process due to its 
much smaller cross section. 
In calculating the production cross sections for $thj$  and $thjb$,
$\mu_F = \mu_R$  are set at $75$ GeV in the four-flavor scheme. 
For $thW$, we are employing the dynamic factorization 
and renormalization scales in the five-flavor scheme.

As shown in Refs.~\cite{higgcision} in which
the model-independent fit to the current Higgs data is performed,
the negative $C^S_t = -1$ is ruled at 95\%CL if only the gauge-Higgs 
coupling $C_v$ and the top-Yukawa coupling $C^S_t$ vary.
However, $C^S_t = -1$ is still allowed at 95\%CL when the gauge-Higgs
$C_v$, top-Yukawa $C^S_t$, bottom-Yukawa $C^S_b$, and tau-Yukawa $C^S_\tau$
couplings are all allowed to vary.
Furthermore, if some sizable contributions 
to $\Delta S^\gamma$ and $\Delta S^g$ due to additional new particles 
running in the loop are assumed, a broad range of $C_t^S$ between $-2$
and $+2$ is still consistent with the current Higgs data.

In the following, we show the results of our numerical analysis in each 
of categories of Leptons ($ss2\ell$ and $3\ell$), $\gamma\gamma$,
and $b\bar{b}$ for the Higgs boson decaying into multileptons, two photons, and
two $b$ quarks, respectively.
Note that, in our numerical analysis, we vary the top-Yukawa coupling $C_t^S$
within the range allowed by the current LHC Higgs data while taking
the SM value for the gauge-Higgs coupling, $C_v=1$. 
For the bottom-Yukawa $C^S_b$ and tau-Yukawa $C^S_\tau$ couplings,
one may freely take either $+1$ or $-1$ since 
their signs would have negligible effects on the production cross sections
and decay branching ratios.
%Note that the negative bottom-Yukawa $C^S_b =-1$ would have negligible
%effects on the production cross section. 

\subsection{Category Leptons for $h \rightarrow$ multileptons}

In the category Leptons
which includes leptonic decays of $h\rightarrow WW, ZZ, \tau\tau\to $
multileptons,
we focus on the subcategories of $ ss2\ell$ and $ 3\ell $ modes. 
We shall use several different values of $ C^{S}_{t} $ to show the 
possibly strong entanglement
between $thX$ production and $t\bar{t}h $ production
for both the ATLAS and CMS analyses.
Note that CMS used 
the so-called Multivariate Analysis (MVA)
method in their analysis, however, 
we only follow their set of preselection cuts and 
event selection requirements to perform the cut-based analysis.

First, we note that the CMS and ATLAS collaborations were adopting different 
signatures and preselection cuts to analyze the category Leptons
as shown in Table~\ref{CMS2-table} and Table~\ref{ATLAS2-table}
\footnote{
This is true also for the $\gamma\gamma$ and $b\bar b$ categories.}.
The CMS analysis was
performed in the $ss2\ell$ and $3\ell$ subcategories
while the ATLAS analysis was carried out in the subcategories of
$2\ell+4j$, $2\ell+\geq 5j$, and $3\ell$.
Without knowing an appropriate way to combine the two sets of data,
we present our results handling
the CMS and ATLAS cases separately to make full use of the existing data.
Further,
in the CMS and ATLAS analyses of the $3\ell$ subcategory, 
also required was
a low-mass invariant-mass cut $M_{\ell\ell} > 12$ GeV 
to remove the $J/\Psi$ background and 
a $Z$-pole mass veto cut $|M_{\ell^{+}\ell^{-}}-M_Z| > 10$ GeV 
to suppress the $Z$ background. 
Some additional cuts on the scalar sum of the transverse momenta ($P_T$) of
the two leptons and the missing energy ($E_T^{\rm miss}$) 
were also applied in the CMS case.

To quantify the effects of different values of $C^{S}_{t} $ on
$t\bar{t}h$ and $thX$, we use the signal-strength formula for
$\mu(t\bar{t}h)$ in Eq.~(\ref{mu-tth}),
which consists of the sum of the products of
the cross section ratios $R$'s and 
the ${\cal D}$-dependent detection efficiency ratio $\epsilon$'s,
which  are in turns 
given by Eq.~(\ref{R-thX}) and Eq.~(\ref{ei}), respectively.
Explicitly, we have
\begin{eqnarray}
 \mu (t\bar{t}h)&=&\frac{\sigma (t\bar{t}h)_{C^{S}_{t}}}{\sigma (t\bar{t}h)_{SM}}\times\epsilon _{1}
+\frac{\sigma (thj)_{C^{S}_{t}}}{\sigma (t\bar{t}h)_{SM}}\times\epsilon _{2}  
+\frac{\sigma (thjb)_{C^{S}_{t}}}{\sigma (t\bar{t}h)_{SM}}\times\epsilon _{3}
+\frac{\sigma (thW)_{C^{S}_{t}}}{\sigma (t\bar{t}h)_{SM}}\times\epsilon _{4} \, 
\nonumber \\[2mm]
 &=& R(t\bar{t}h)\times\epsilon _{1}+R(thj)\times\epsilon _{2}+R(thjb)\times\epsilon _{3}+R(thW)\times\epsilon _{4} \,.
\label{mu}
\end{eqnarray}

\begin{table}[t!]
\caption{\small \label{R(thx)}
The cross-section ratios
$R(t\bar t h)$ and $R(thX)$ with $X=j,jb,W$ defined in Eq.~(\ref{R-thX}).
We are taking $\sqrt{s}=8$ TeV (LHC-8) and
$C_t^S = \pm 1\,, \pm 1.5$.
}
\medskip
\begin{ruledtabular}
\begin{tabular}{lcccc}
LHC-8  & $C^{S}_{t}=1$ & $C^{S}_{t}=-1$ & $C^{S}_{t}=1.5$ & $C^{S}_{t}=-1.5$ \\
\hline
Cross Section of $t\bar{t} h$(pb) &0.13&\\%0.13&0.30&0.30\\
$ R(t\bar{t}h) $ & 1 & 1 &2.25 &2.25 \\ 
$ R(thj) $ & 8.36e-2 & 1.08 & 0.15 & 1.66 \\
$ R(thjb) $ & 4.30e-2 & 0.54 & 8.56e-2 &   0.84 \\
$ R(thW) $ & 3.21e-2&0.19 &7.05e-2 &0.31 \\
\end{tabular}
\end{ruledtabular}
\end{table}
In Table~\ref{R(thx)}, we show the cross section ratios
$R(t\bar t h)$ and $R(thX)$ with $X=j,jb,W$ at the 8 TeV LHC (LHC-8)
taking $C_t^S = \pm 1$ and $\pm 1.5$. Note $R(t\bar t h)=1\,(2.25)$ for 
$|C_t^S| = 1\,(1.5)$ and the $thX$ cross sections can be largely enhanced
for the negative values of $C_t^S$.

In Table~\ref{sum1-CMS}, we show
the ${\cal D}$-dependent detection efficiency ratios $\epsilon_{1,2,3,4}$
with the CMS cuts
in the $ss2\ell$ (upper) and $3\ell$ (lower) subcategories for 
$C_t^S = \pm 1\,,\pm 1.5$.
By using the cross section ratios given in Table~\ref{R(thx)}, 
one can obtain the CMS $t\bar t h$ signal strengths $\mu_{tth}^{\rm CMS}$.
We observe that
$\mu_{tth\,,ss2\ell}^{\rm CMS} \sim 2\,(3)$ for $C_t^S =1.5\,(-1.5)$
and the signal strengths are larger for the negative values of $C_t^S$.
One may make similar observations for $\mu_{tth\,,3\ell}^{\rm CMS}$.
Recently, the CMS collaboration has also reported a
possible excess in the decay process $h\rightarrow\tau^{\mp}\mu^{\pm}$,
$B(h\rightarrow\tau^{\mp}\mu^{\pm})=0.84\,^{+0.39}_{-0.37}\%$,
with a significance of $ 2.4\sigma$ in the search for the lepton-flavor 
violation (LFV)\cite{Khachatryan:2015kon}. 
If we take into account this LFV decay of the Higgs boson, 
we can slightly enhance the production rate of  
$h \rightarrow$ multileptons mode by a few percents.
We estimate the $h\rightarrow \tau^\mp\mu^\pm$ 
contribution by rescaling $h\rightarrow \tau^+\tau^-$ channel with 
the branching ratios and the $\tau$ detection efficiency.
The CMS $t\bar t h$ signal strengths $\mu_{tth}^{\rm CMS}$ after taking account
of $h\rightarrow\tau^{\mp}\mu^{\pm}$ are also presented in Table~\ref{sum1-CMS}.

\begin{table}[t!]
%\begin{table}[th!]
\caption{\small \label{sum1-CMS}
Category Leptons:
The ${\cal D}$-dependent detection efficiency ratios $\epsilon_i$'s
defined in Eq.~(\ref{ei}) with the CMS cuts for
the category Leptons taking
$C_t^S=\pm1\,,\pm 1.5$ and $\sqrt{s} = 8$ TeV.
The resulting signal strengths
$\mu_{tth\,,ss2\ell}^{\rm CMS}$ and
$\mu_{tth\,,3\ell}^{\rm CMS}$
are also shown.
The last row in each partition shows the values of 
$\mu_{tth\,,ss2\ell}^{\rm CMS}$ and
$\mu_{tth\,,3\ell}^{\rm CMS}$
including the contributions from $h \to \mu^\pm \tau^\mp$.
}
\medskip
\begin{ruledtabular}
\begin{tabular}{lcccc}
 LHC-8 & \multicolumn{4}{c}{With CMS Analysis Cuts} \\
\hline 
  & $C^{S}_{t}=1$ & $C^{S}_{t}=-1$ & $C^{S}_{t}=1.5$ & $C^{S}_{t}=-1.5$ \\
\hline 
& \multicolumn{4}{c}{The category of $ ss2\ell $} \\
\hline 
Efficiency of $t\bar{t} h$ &9.02e-4&\\%8.53e-4&8.87e-4&8.78e-4\\
$ \epsilon _{1} $ & 1 & 0.95 & 0.98&0.97 \\
$ \epsilon _{2} $ &0.1 &0.12 &0.12 &0.13 \\
$ \epsilon _{3} $ &0.35 &0.38 &0.33 &0.39 \\
$ \epsilon _{4} $ & 0.68& 0.85& 0.72&0.83 \\
$\mu_{tth\,,ss2\ell}^{\rm CMS}$ &1.05 &1.45 &2.31 &2.99 \\
$\mu_{tth\,,ss2\ell}^{\rm CMS}$
including $h\rightarrow \tau^\mp \mu^\pm$ &1.09&1.51 &2.40 &3.11 \\
\hline\hline
  & \multicolumn{4}{c}{The category of $ 3\ell $} \\
\hline 
Efficiency of $t\bar{t} h$ &9.54e-4&\\%9.05e-4&9.52e-4&9.25e-4\\
$ \epsilon _{1} $ & 1 & 0.95 & 1&0.97 \\
$ \epsilon _{2} $ & 0.34& 0.40& 0.37& 0.42\\
$ \epsilon _{3} $ &0.55&0.61 & 0.57&0.65 \\
$ \epsilon _{4} $ & 0.90& 1.14& 0.98&1.13 \\
$\mu_{tth\,,3\ell}^{\rm CMS}$ &1.08&1.93 &2.42 &3.77 \\
$\mu_{tth\,,3\ell}^{\rm CMS}$
including $h\rightarrow \tau^\mp \mu^\pm$ &1.12&2.01 &2.51 &3.92\\
\end{tabular}
\end{ruledtabular}
\end{table}

Similarly, we calculate the ${\cal D}$-dependent detection efficiency ratios
$\epsilon_{1,2,3,4}$ with the ATLAS cuts in the
$2\ell+4j$ (upper), $2\ell+\geq 5j$ (middle), and $3\ell$ (lower) subcategories
for several values of $C_t^S$ and present them in Table~\ref{sum1.1-ATLAS},
together with the ATLAS $t\bar t h$ signal strengths $\mu_{tth}^{\rm ATLAS}$.
Similar observations can be made as in the CMS case.

%\begin{table}[th!]
\begin{table}[t!]
\caption{\small \label{sum1.1-ATLAS}
Category Leptons:
The same as Table~\ref{sum1-CMS} but with the ATLAS cuts.
}
\medskip
\begin{ruledtabular}
\begin{tabular}{lcccc}
  LHC-8& \multicolumn{4}{c}{With ATLAS Analysis Cuts} \\
\hline 
  & $C^{S}_{t}=1$ & $C^{S}_{t}=-1$ & $C^{S}_{t}=1.5$ & $C^{S}_{t}=-1.5$ \\
\hline
 & \multicolumn{4}{c}{The category of $ 2\ell+4j $} \\
\hline
Efficiency of $t\bar{t} h$ &4.27e-4&\\%4.49e-4&4.10e-4&4.28e-4\\
$ \epsilon _{1} $ &1&1.05&0.96&1.0\\
$ \epsilon _{2} $ &0.31&0.31&0.32&0.38
\\
$ \epsilon _{3} $ &0.52&0.63&0.53&0.57 \\
$ \epsilon _{4} $ & 1.10&1.16&1.09&1.18 \\
$\mu_{tth\,,2\ell+4j}^{\rm ATLAS}$
& 1.08&1.96&2.33&3.72\\
$\mu_{tth\,,2\ell+4j}^{\rm ATLAS}$
including $h\rightarrow \tau^\mp \mu^\pm$ &1.13&2.03 &2.42 &3.86 \\

\hline
  & \multicolumn{4}{c}{The category of $ 2\ell+\geq 5j $} \\
\hline 
Efficiency of $t\bar{t} h$ &5.25e-4&\\%4.84e-4&4.78e-4&4.86e-4\\
$ \epsilon _{1} $ & 1&0.92&0.91&0.93\\
$ \epsilon _{2} $ & 0.08&0.08&0.09&0.09
\\
$ \epsilon _{3} $ &0.25&0.28&0.22&0.26 \\
$ \epsilon _{4} $ & 0.74&0.94&0.75&0.95\\
$\mu_{tth\,,2\ell+\geq 5j}^{\rm ATLAS}$
&1.04&1.35&2.13&2.74 \\
$\mu_{tth\,,2\ell+\geq 5j}^{\rm ATLAS}$
including $h\rightarrow \tau^\mp \mu^\pm$ &1.08&1.40 &2.22 &2.85 \\
\hline
  & \multicolumn{4}{c}{The category of $ 3\ell $} \\
\hline 
Efficiency of $t\bar{t} h$ &1.05e-4&\\%9.39e-5&8.72e-5&9.49e-5\\
$ \epsilon _{1} $ &1&0.89&0.83&0.90 \\
$ \epsilon _{2} $ &0.06&0.09&0.13&0.09 
\\
$ \epsilon _{3} $ & 0.34&0.45&0.30&0.47\\
$ \epsilon _{4} $ &  0.89&1.5&0.92&1.61\\
$\mu_{tth\,,3\ell}^{\rm ATLAS}$
&  1.05&1.52&1.97&3.07\\
$\mu_{tth\,,3\ell}^{\rm ATLAS}$
including $h\rightarrow \tau^\mp \mu^\pm$ &1.09&1.58 &2.05 &3.19 \\
\end{tabular}
\end{ruledtabular}
\end{table}

\begin{figure}[t!]
\centering
\includegraphics[width=3.in]{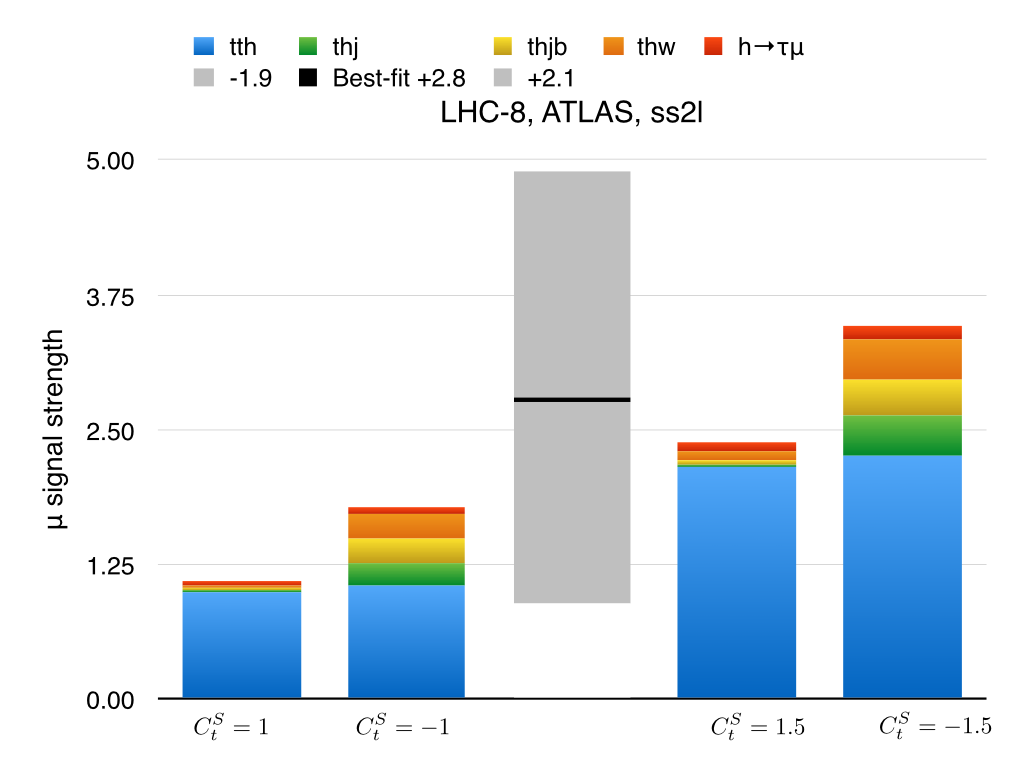}
\includegraphics[width=3.in]{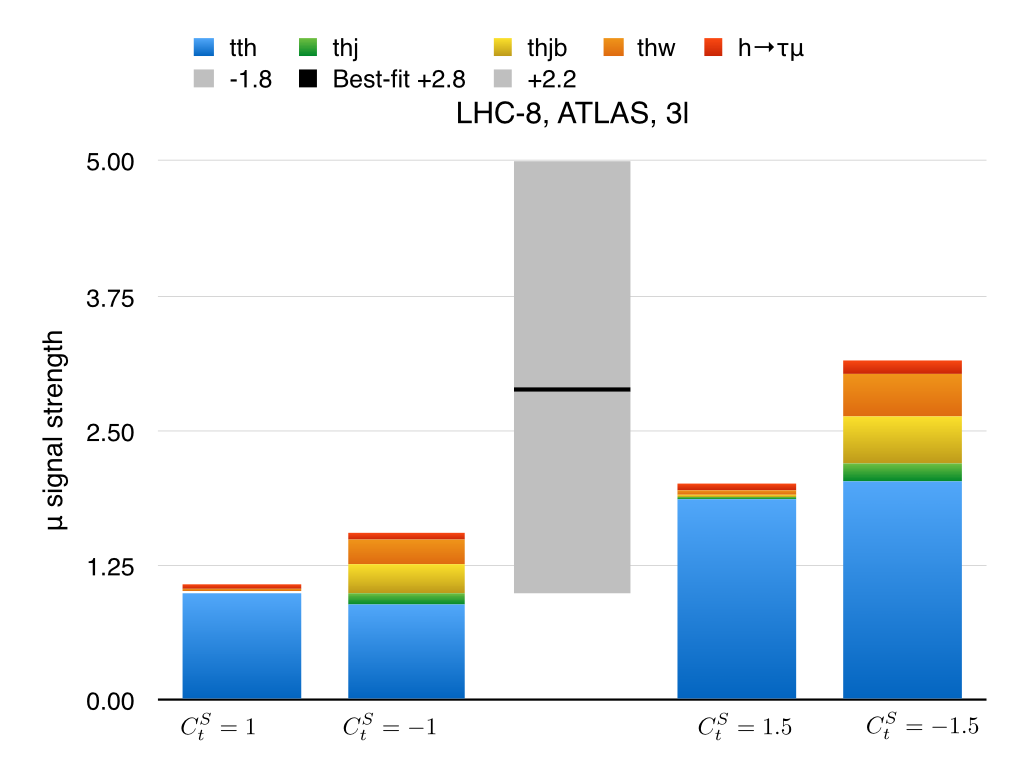}
\includegraphics[width=3.in]{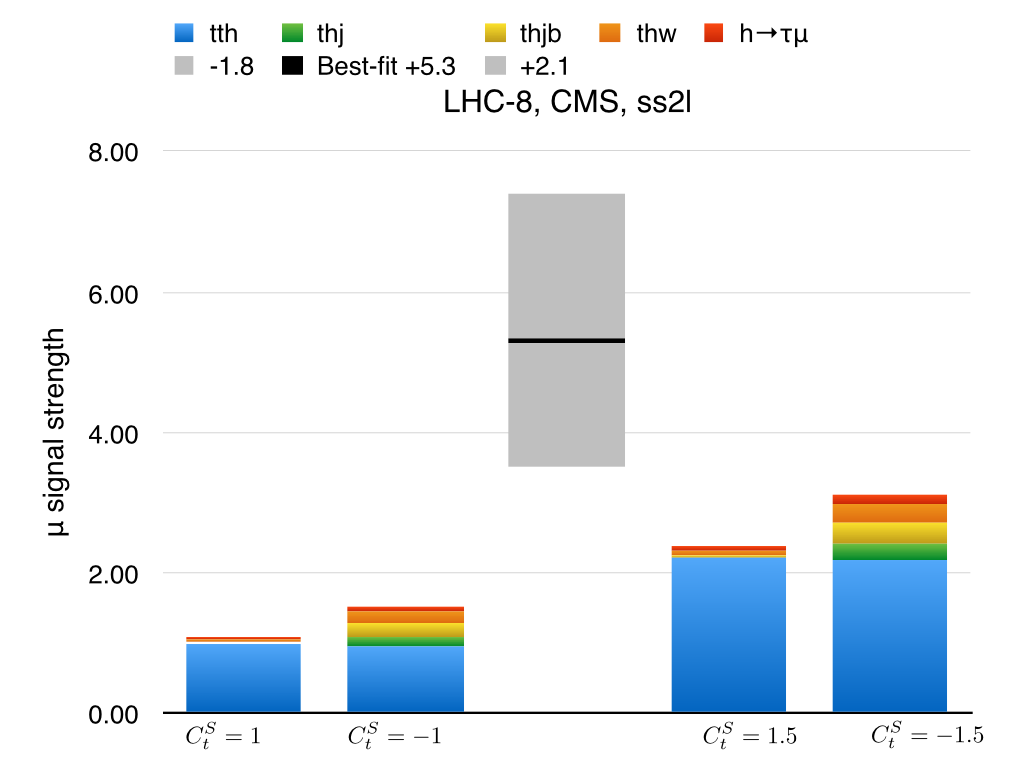}
\includegraphics[width=3.in]{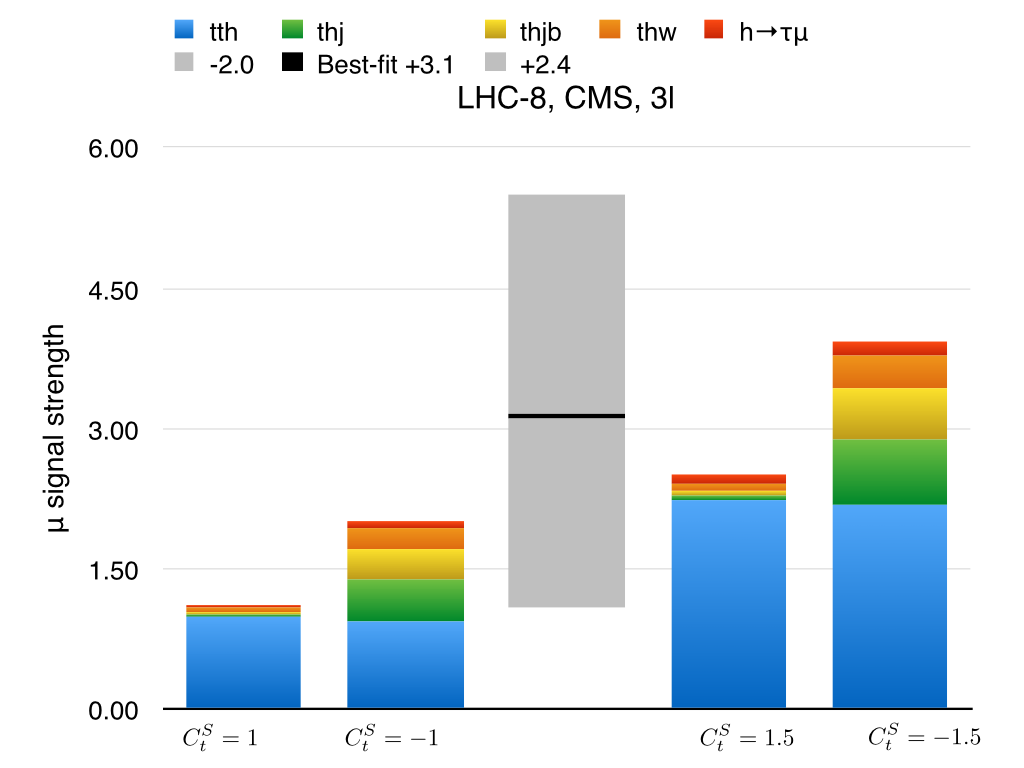}
\caption{\small \label{sum-1-atlas-cms}
Category Leptons:
Accumulated signal strengths 
$\mu_{tth\,,ss2\ell}^{\rm ATLAS}$ (upper left),
$\mu_{tth\,,3\ell}^{\rm ATLAS}$ (upper right),
$\mu_{tth\,,ss2\ell}^{\rm CMS}$ (lower left), and
$\mu_{tth\,,3\ell}^{\rm CMS}$ (lower right) at LHC-8
obtained by stacking the various $thX$ contributions 
on the $t\bar t h$ one
for $C_t^S=+1\,,-1\,,+1.5\,,-1.5$ from left to right.
The grey columns in the center represent  the current 
8 TeV LHC data from Table~\ref{CMS-ATLAS1-table}.
}
\end{figure}

Finally, we show in Fig.~\ref{sum-1-atlas-cms} the accumulative signal
strengths
$\mu_{tth\,,ss2\ell}^{\rm ATLAS}$ (upper left),
$\mu_{tth\,,3\ell}^{\rm ATLAS}$ (upper right),
$\mu_{tth\,,ss2\ell}^{\rm CMS}$ (lower left), and
$\mu_{tth\,,3\ell}^{\rm CMS}$ (lower right) at $\sqrt{s}=8$ TeV
by stacking the various $thX$ contributions 
on the $t\bar t h$ one
for $C_t^S=+1\,,-1\,,+1.5\,,-1.5$ from left to right.
The grey columns in the center without $C_t^S$ value represent  the current 
8 TeV LHC data, see Table~\ref{CMS-ATLAS1-table}.  
The ATLAS $ss2\ell$ signal strength
$\mu_{tth\,,ss2\ell}^{\rm ATLAS}$ is 
obtained by counting the event rates by combining
the $2\ell+4j$ and $2\ell+\geq 5j$ selections.

%
%\begin{figure}[th!]

%%% TABLES for Category $\gamma\gamma$
%\begin{table}[t!]
\begin{table}[th!]
\caption{\small \label{sumrr-CMS}
Category $\gamma\gamma$:
The ${\cal D}$-dependent detection efficiency ratios $\epsilon_i$'s
defined in Eq.~(\ref{ei}) with the CMS cuts for
the category $\gamma\gamma$ 
taking $C_t^S=\pm 1\,,\pm 1.5$ and $\sqrt{s} = 8$ TeV.
The resulting signal strengths
$\mu_{tth\,,\gamma\gamma (lep)}^{\rm CMS}$ and
$\mu_{tth\,,\gamma\gamma (had)}^{\rm CMS}$
are also shown.
Note that the CMS cuts $N(j)\ge4, N(b)\ge 1$ for 
$\gamma\gamma$ hadronic channel are not strong enough to separate 
$t\bar{t} h$ from $thX$ processes.
}
\medskip
\begin{ruledtabular}
\begin{tabular}{lcccc}
  LHC-8& \multicolumn{4}{c}{With CMS Analysis Cuts} \\
  \hline 
  & $C^{S}_{t}=1$ & $C^{S}_{t}=-1$ & $C^{S}_{t}=1.5$ & $C^{S}_{t}=-1.5$ \\
\hline 
 & \multicolumn{4}{c}{Leptonic Selection} \\
\hline
Efficiency of $t\bar{t} h$ &1.13e-5&\\%2.16e-5&7.55e-6&2.61e-5\\
$ \epsilon _{1} $ &1&{0.81}&{0.99}&{0.92}\\
$ \epsilon _{2} $ &0.03&{0.02}&{0.01}&{0.02}
\\
$ \epsilon _{3} $ &0.11&{0.10}&{0.08}&{0.10} \\
$ \epsilon _{4} $ & 0.57&{0.86}&{0.83}&{0.76} \\
$\mu_{tth\,,\gamma\gamma (lep)}^{\rm CMS}$
& 1.03&1.05&2.29&2.42\\
\hline 
 & \multicolumn{4}{c}{Hadronic Selection}\\
\hline
Efficiency of $t\bar{t} h$ &1.47e-4&\\%3.43e-4&1.03e-4&3.57e-4\\
$ \epsilon _{1} $ &1&{0.99}&{1.05}&{0.97}\\
$ \epsilon _{2} $ &0.44&{0.47}&{0.43}&{0.50}
\\
$ \epsilon _{3} $ &1.40&{1.62}&{1.47}&{1.58}\\
$ \epsilon _{4} $ & 0.52&{0.65}&{0.54}&{0.68} \\
$\mu_{tth\,,\gamma\gamma (had)}^{\rm CMS}$
& 1.11&2.50&2.59&4.56\\
\hline\hline
Combined $\mu_{tth\,,\gamma\gamma}^{\rm CMS}$
&1.11&2.40&2.57&4.41\\
\end{tabular}
\end{ruledtabular}
\end{table}
%
%
%\begin{table}[th!]
\begin{table}[t!]
\caption{\small \label{sumrr-ATLAS}
Category $\gamma\gamma$:
The same as Table~\ref{sumrr-CMS} with the ATLAS cuts.
}
\medskip
\begin{ruledtabular}
\begin{tabular}{lcccc}
  LHC-8& \multicolumn{4}{c}{With ATLAS Analysis Cuts} \\
  \hline 
  & $C^{S}_{t}=1$ & $C^{S}_{t}=-1$ & $C^{S}_{t}=1.5$ & $C^{S}_{t}=-1.5$ \\
\hline 
 & \multicolumn{4}{c}{Leptonic Selection } \\
\hline
Efficiency of $t\bar{t} h$ &8.15e-6&\\%1.34e-5&5.26e-6&2.44e-5\\
$ \epsilon _{1} $ &1.00&{0.69}&{0.96}&{1.19}\\
$ \epsilon _{2} $ &0.03&{0.08}&{0.11}&{0.06}\\
$ \epsilon _{3} $ &0.14&{0.11}&{0.06}&{0.11} \\
$ \epsilon _{4} $ & 0.82&{0.74}&{1.06}&{0.43} \\
$\mu_{tth\,,\gamma\gamma (lep)}^{\rm ATLAS}$
& 1.03&0.99&2.25&3.01\\

\hline 
 & \multicolumn{4}{c}{Hadronic Selection } \\
\hline
Efficiency of $t\bar{t} h$ &1.06e-4&\\%2.52e-4&7.34e-5&2.61e-4\\
$ \epsilon _{1} $ &1&{1.01}&{1.03}&{0.98}\\
$ \epsilon _{2} $ &0.05&{0.05}&{0.05}&{0.07}
\\
$ \epsilon _{3} $ &0.39&{0.46}&{0.44}&{0.46} \\
$ \epsilon _{4} $ & 0.39&{0.49}&{0.38}&{0.48} \\
$\mu_{tth\,,\gamma\gamma (had)}^{\rm ATLAS}$
& 1.03&1.40&2.39&2.85\\
\hline\hline
Combined 
$\mu_{tth\,,\gamma\gamma}^{\rm ATLAS}$
&1.03&1.37&2.38&2.86\\
\end{tabular}
\end{ruledtabular}
\end{table}
\subsection{ Category $\gamma\gamma$ for
$h \rightarrow$ $\gamma\gamma$}

In the  category $\gamma\gamma$ for
$h \rightarrow$ $\gamma\gamma$,
we include all the decay modes of the top-quark pair.
We consider two subcategories of {\it leptonic selection} 
and {\it hadronic selection}.
The lepton-selection subcategory is for the
semileptonically and leptonically decaying top-quark pair
while the hadronic-selection one for
the hadronically decaying top-quark pair.
To single out the effect of anomalous top-Yukawa coupling on
$thj$ and $tth$ production in this category, 
we assume a non-vanishing $\Delta S^\gamma$ due to 
additional particles 
running in the $h$-$\gamma$-$\gamma$ loop, see Eq.~(\ref{eq:hrr}).
In fact, one may have
$B(h\rightarrow \gamma\gamma) = (2.3, 5.4, 1.53, 5.68)\times
10^{-3}$ 
for $C^S_t=(1,\ -1,\ 1.5,\ -1.5)$ using, for example, 
{\tt HDECAY} \cite{hdecay}.
We are using $B(h\rightarrow \gamma\gamma) = 2.3\times 10^{-3}$ independently
of $C_t^S$ assuming a non-zero $\Delta S^\gamma$ which cancels out
the the effect of anomalous top-Yukawa coupling on $B(h\rightarrow
\gamma\gamma)$.
This assumption also helps to avoid
the constraint on $S^\gamma(m_h)$
from the current LHC Higgs data~\cite{higgcision}.

To repeat the CMS analysis, we follow their selection cuts 
listed in Table~\ref{CMS2-table}, which are used in the cut-based analysis
\cite{Khachatryan:2014qaa}.
Also, we further impose the Higgs-mass window cut: 
$100\,{\rm GeV}  \leq m_{\gamma\gamma}\leq 180$ GeV. 

For the ATLAS analysis, we follow Ref.~\cite{Aad:2014lma} 
with preselection cuts listed in
Table~\ref{ATLAS2-table}. We further impose the Higgs-mass window
cut ($105 \,{\rm GeV} < m_{\gamma\gamma} < 160\,{\rm GeV}$) and 
the $\Delta R$ cuts: 
$\Delta R_{l\gamma} > 0.4$, $\Delta R_{j\gamma} > 0.4$, $\Delta R_{j\mu} >
0.4$, $\Delta R_{j e} > 0.2$. 
The missing energy cut $E_T^{miss}>20$ GeV and the $e \gamma$ invariant-mass 
cut $M_{e\gamma}> 94 \, {\rm GeV}$ or $ < 84\, {\rm GeV}$ 
are also applied in the leptonic-selection category.  
In the hadronic-selection subcategory, we adopt the
selection 1 in Ref.~\cite{Aad:2014lma} 
using the working point with efficiency of $70\%$
for identifying $b$-jets.

%\begin{figure}[th!]
\begin{figure}[t!]
\centering
\includegraphics[width=3in]{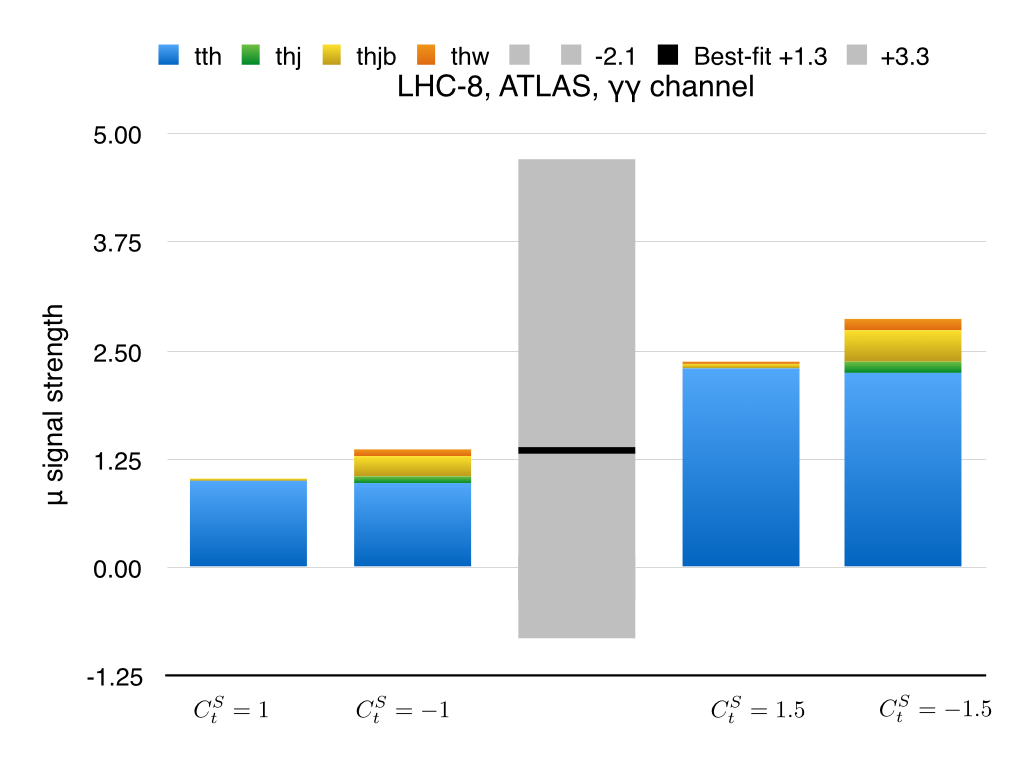}
\includegraphics[width=3in]{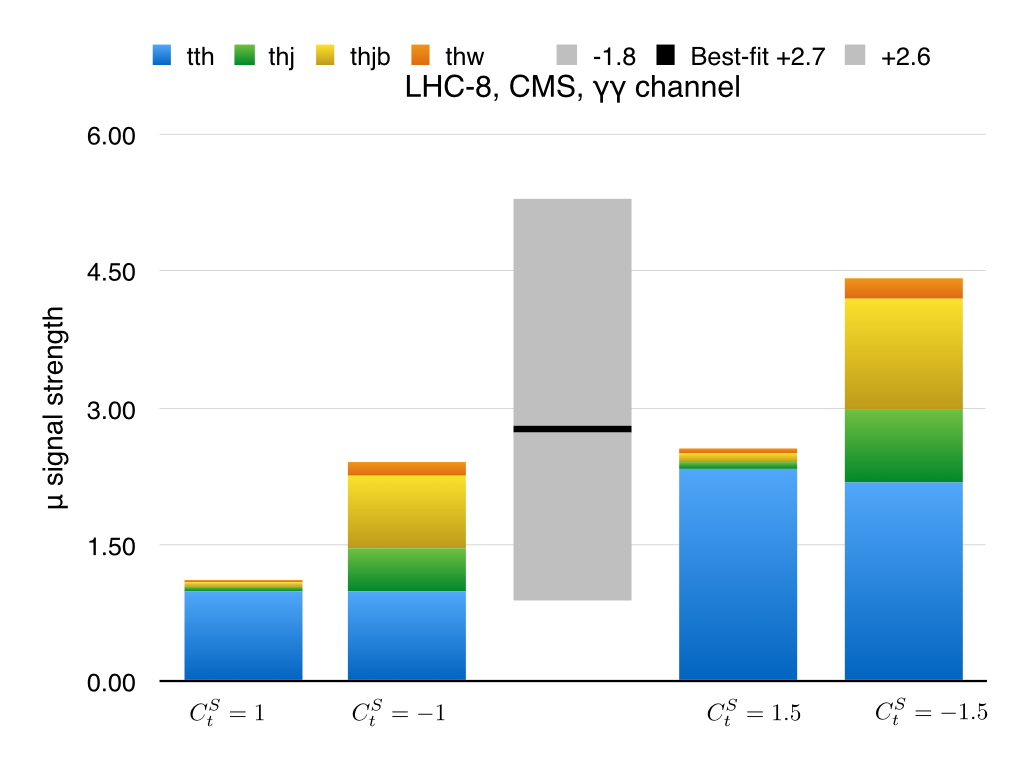}
\caption{\small \label{sum-rr-atlas-cms}
Category $\gamma\gamma$:
Accumulated signal strengths
$\mu_{tth\,,\gamma\gamma}^{\rm ATLAS}$ (left) and
$\mu_{tth\,,\gamma\gamma}^{\rm CMS}$ (lower right) at LHC-8
obtained by stacking the various $thX$ contributions
on the $t\bar t h$ one
for $C_t^S=+1\,,-1\,,+1.5\,,-1.5$ from left to right.
The grey columns in the center represent  the current
8 TeV LHC data from Table~\ref{CMS-ATLAS1-table}.}
\end{figure}

Before we present the results of our numerical study of
the effects of $thX$ on $t\bar{t}h$ in the category $\gamma\gamma$, 
we would like to
make some remarks on a few noticeable aspects from
the ATLAS $t\bar th$ search. 
It has been shown that there was no significant excess over the background 
in the $h \to \gamma\gamma $ mode, and thus 
the 95\% CL upper limit is set at $6.7 \times \sigma_{\rm SM}(t\bar{t}h)$.
Especially, ATLAS
took into account the dependence of 
the $t\bar{t}h$ and $thX$ cross sections as well as the branching ratio
$B(h \rightarrow\gamma\gamma)$ on the top-Yukawa coupling. 
The ATLAS $t\bar th$ search sets  the lower and upper limits on $C_t^S$:
$-1.3\leq C_t^S \leq 8.0$ at 95\% CL.

In Table~\ref{sumrr-CMS}, we show
the ${\cal D}$-dependent detection efficiency ratios $\epsilon_{1,2,3,4}$
with the CMS cuts
in the hadronic-selection (upper) and leptonic-selection (lower) subcategories 
for $C_t^S = \pm 1\,, \pm 1.5$.
By using the cross section ratios given in Table~\ref{R(thx)}, 
one can obtain the CMS $t\bar t h$ signal strengths 
$\mu_{tth\,,\gamma\gamma (lep)}^{\rm CMS}$  and
$\mu_{tth\,,\gamma\gamma (had)}^{\rm CMS}$ using Eq.~(\ref{mu}).
In the leptonic-selection subcategory, 
we observe that 
$\mu_{tth\,,\gamma\gamma (lep)}^{\rm CMS}>2$ for $|C_t^S| =1.5$. 
In the hadronic-selection subcategory, we obtain the larger values 
for negative $C_t^S$:
$\mu_{tth\,,\gamma\gamma (had)}^{\rm CMS}\sim 1,\  2.5;\ 2.5 ,\ 5$
for $C^S_t=(+1,\ -1;\ +1.5\ -1.5)$.
Also presented is the combined signal strength 
$\mu_{tth\,,\gamma\gamma}^{\rm CMS}$ which is
obtained by counting the event rates by combining the hadronic
and leptonic selections. 
Similarly, in  Table~\ref{sumrr-ATLAS}, 
we show the ${\cal D}$-dependent detection efficiency ratios $\epsilon_{1,2,3,4}$
with the ATLAS cuts and the signal strengths 
$\mu_{tth\,,\gamma\gamma (lep)}^{\rm ATLAS}$, 
$\mu_{tth\,,\gamma\gamma (had)}^{\rm ATLAS}$,  and
$\mu_{tth\,,\gamma\gamma}^{\rm ATLAS}$.
Similar observations can be made as in the CMS case.

Finally, in Fig.~\ref{sum-rr-atlas-cms}, we show the accumulative 
combined signal strengths
$\mu_{tth\,,\gamma\gamma}^{\rm ATLAS}$ (left) and
$\mu_{tth\,,\gamma\gamma}^{\rm CMS}$ (right) at $\sqrt{s}=8$ TeV
by stacking the various $thX$ contributions
on the $t\bar t h$ one
for $C_t^S=+1\,,-1\,,+1.5\,,-1.5$ from left to right.
The grey columns in the center without $C_t^S$ value represent  the current
8 TeV LHC data, see Table~\ref{CMS-ATLAS1-table}.

%%% TABLES for Category $b\bar b$
%\begin{table}[th!]
\begin{table}[t!]
\caption{\small \label{sumbb-CMS}
Category $b\bar b$:
The ${\cal D}$-dependent detection efficiency ratios $\epsilon_i$'s
defined in Eq.~(\ref{ei}) with the CMS cuts for
the category $b\bar b$ 
taking $C_t^S=\pm 1\,,\pm 1.5$ and $\sqrt{s} = 8$ TeV.
The resulting signal strengths
$\mu_{tth\,,b\bar b (1\ell)}^{\rm CMS}$ and
$\mu_{tth\,,b\bar b (2\ell)}^{\rm CMS}$ and
the combined one 
$\mu_{tth\,,b\bar b}^{\rm CMS}$
are also shown.
}
\medskip
\begin{ruledtabular}
\begin{tabular}{lcccc}
  LHC-8& \multicolumn{4}{c}{With CMS Analysis Cuts} \\
  \hline 
  & $C^{S}_{t}=1$ & $C^{S}_{t}=-1$ & $C^{S}_{t}=1.5$ & $C^{S}_{t}=-1.5$ \\
\hline 
 & \multicolumn{4}{c}{Single Lepton } \\
\hline
Efficiency of $t\bar{t} h$ & 1.17e-1&\\% 1.17e-1& 1.15e-1& 1.16e-1\\
$ \epsilon _{1} $ &1&1&0.98&1\\
$ \epsilon _{2} $ &0.39&0.42&0.38&0.41
\\
$ \epsilon _{3} $ & 0.67&0.69&0.69&0.70\\
$ \epsilon _{4} $ &  0.81&0.89&0.82&0.89\\
$\mu_{tth\,,b\bar b (1\ell)}^{\rm CMS}$
& 1.09&2.01&2.38&3.79\\
\hline 
 & \multicolumn{4}{c}{Dilepton } \\
\hline
Efficiency of $t\bar{t} h$ &2.03e-2&\\%2.22e-2&2.11e-2&2.02e-2\\
$ \epsilon _{1} $ &1&1.09&1.04&0.99\\
$ \epsilon _{2} $ &0.14&0.18&0.17&0.17
\\
$ \epsilon _{3} $ & 0.33&0.37&0.33&0.36\\
$ \epsilon _{4} $ & 0.73&0.87&0.77&0.80 \\
$\mu_{tth\,,b\bar b (2\ell)}^{\rm CMS}$
&1.05&1.66&2.45&3.05 \\
\hline\hline
Combined 
$\mu_{tth\,,b\bar b}^{\rm CMS}$
&1.08&1.96&2.39&3.68\\
\end{tabular}
\end{ruledtabular}
\end{table}
%
%\begin{table}[th!]
\begin{table}[t!]
\caption{\small \label{sumbb-ATLAS}
Category $b\bar b$:
The same as Table~\ref{sumbb-CMS} but with the ATLAS cuts.
}
\medskip
\begin{ruledtabular}
\begin{tabular}{lcccc}
  LHC-8& \multicolumn{4}{c}{With ATLAS Analysis Cuts} \\
  \hline 
  & $C^{S}_{t}=1$ & $C^{S}_{t}=-1$ & $C^{S}_{t}=1.5$ & $C^{S}_{t}=-1.5$ \\
\hline 
 & \multicolumn{4}{c}{Single Lepton } \\
\hline
Efficiency of $t\bar{t} h$ & 1.19e-1&\\% 1.21e-1& 1.18e-1& 1.18e-1\\
$ \epsilon _{1} $ &1&1.01&0.99&0.99\\
$ \epsilon _{2} $ &0.36&0.40&0.40&0.42
\\
$ \epsilon _{3} $ & 0.66&0.68&0.69&0.69\\
$ \epsilon _{4} $ &  0.78&0.88&0.80&0.88\\
$\mu_{tth\,,b\bar b (1\ell)}^{\rm ATLAS}$
& 1.08&1.98&2.40&3.78\\
\hline 
 & \multicolumn{4}{c}{Dilepton } \\
\hline
Efficiency of $t\bar{t} h$ &1.57e-2&\\%1.61e-2&1.68e-2&1.49e-2\\
$ \epsilon _{1} $ &1&1.02&1.07&0.95\\
$ \epsilon _{2} $ &0.07&0.08&0.06&0.08
\\
$ \epsilon _{3} $ & 0.11&0.13&0.11&0.11\\
$ \epsilon _{4} $ & 0.86&0.86&0.86&0.86 \\
$\mu_{tth\,,b\bar b (2\ell)}^{\rm ATLAS}$
&1.04&1.35&2.50&2.63 \\
\hline\hline
Combined 
$\mu_{tth\,,b\bar b}^{\rm ATLAS}$
&1.08&1.91&2.41&3.64\\
\end{tabular}
\end{ruledtabular}
\end{table}
\subsection{Category $b\bar b$ for $h \rightarrow$ $b\bar{b}$}

In the category $b\bar b$ for $h \rightarrow$ $b\bar{b}$, we consider the
semileptonic and leptonic decay modes of the top-quark pair
which leads to the two subcategories of {\it single lepton} ($1\ell$) and
{\it dilepton} ($2\ell$).
The CMS preselection cuts shown in Table~\ref{CMS2-table} and
the ATLAS ones in Table~\ref{ATLAS2-table} are first applied.
And we further impose $P_{T_j}> 40$ GeV for
the leading 3 jets in the single-lepton subcategory.
In the dilepton subcategory, we select
the events with exactly two oppositely charged leptons 
$e^+ e^- , e^\pm \mu^\mp, \mu^+\mu^-$
with $P_{T_{l_1}} > 25$ GeV and $P_{T_{l_2}} > 15$ GeV.
For $e^\pm\mu^\mp$ events, we further require $H_T$,
scalar sum of transverse momenta of leptons and jets,
to be larger than $130$ GeV. 
For $e^+ e^-$ and $\mu^+ \mu^-$ events, we impose two more conditions: 
{\it (i)} more than 2 $b$-jets and $M_{ll}>15$ GeV to 
reduce the $J/\Psi $ background and  
{\it (ii)} exactly 2 $b$-jets, $M_{ll}>60$ GeV to remove the 
events in the low-mass region with large error bars, and
$|M_{ll}-M_Z|>8$ GeV to veto the $Z$ background. 
We then combine these selections to complete the dilepton selection.

In Table~\ref{sumbb-CMS}, we show
the ${\cal D}$-dependent detection efficiency ratios $\epsilon_{1,2,3,4}$
with the CMS cuts
in the single-lepton (upper) and dilepton (lower) subcategories
for $C_t^S = \pm 1\,, \pm 1.5$.
By using the cross section ratios given in Table~\ref{R(thx)},
one can obtain the CMS $t\bar t h$ signal strengths
$\mu_{tth\,,b\bar b (1\ell)}^{\rm CMS}$  and
$\mu_{tth\,,b\bar b (2\ell)}^{\rm CMS}$ using Eq.~(\ref{mu}).
We observe that
$\mu_{tth\,,b\bar b (1\ell)}^{\rm CMS}>2$  for $C_t^S=-1\,,\pm 1.5$ and
$\mu_{tth\,,b\bar b (2\ell)}^{\rm CMS}>2$  for $C_t^S=\pm 1.5$.
The combined signal strength $\mu_{tth\,,b\bar b}^{\rm CMS}\gsim 2$
for $C_t^S =-1\,,\pm 1.5$. 

Similarly, in  Table~\ref{sumbb-ATLAS},
we show the ${\cal D}$-dependent detection efficiency ratios
$\epsilon_{1,2,3,4}$
with the ATLAS cuts and the signal strengths
$\mu_{tth\,,b\bar b (1\ell)}^{\rm ATLAS}$,
$\mu_{tth\,,b\bar b (2\ell)}^{\rm ATLAS}$,  and
$\mu_{tth\,,b\bar b}^{\rm ATLAS}$.
Similar observations can be made as in the CMS case.

\begin{figure}[t!]
%\begin{figure}[th!]
%\begin{figure}[h!]
\centering
\includegraphics[width=3in]{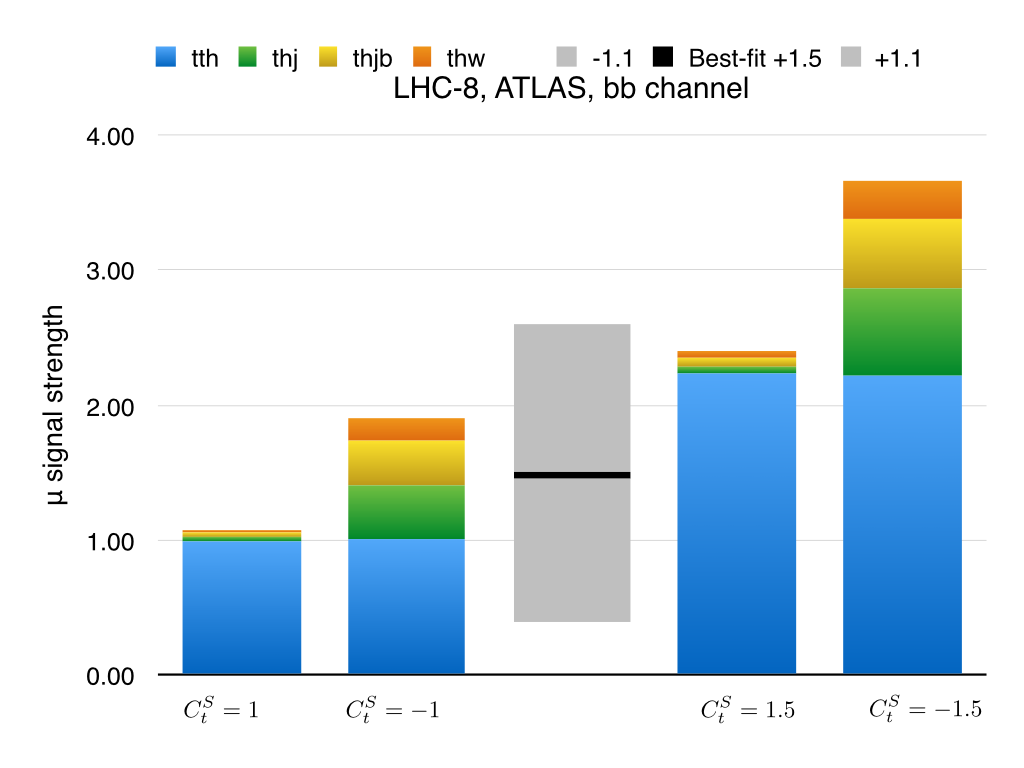}
\includegraphics[width=3in]{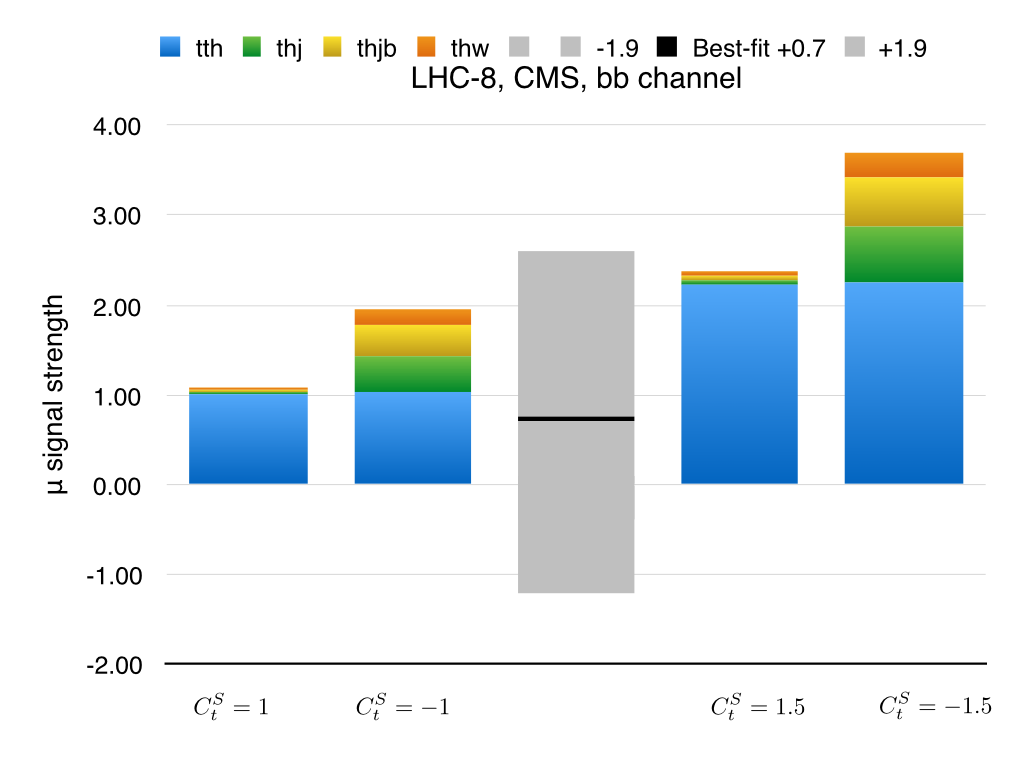}
\caption{\small \label{sum-bb-atlas-cms}
Category $b\bar b$:
Accumulated signal strengths 
$\mu_{tth\,,b\bar b}^{\rm ATLAS}$ (left) and
$\mu_{tth\,,b\bar b}^{\rm CMS}$ (lower right) at LHC-8
obtained by stacking the various $thX$ contributions 
on the $t\bar t h$ one
for $C_t^S=+1\,,-1\,,+1.5\,,-1.5$ from left to right.
The grey columns in the center represent  the current 
8 TeV LHC data from Table~\ref{CMS-ATLAS1-table}.
}
\end{figure}

Finally, we show in Fig.~\ref{sum-bb-atlas-cms} the accumulative
combined signal strengths
$\mu_{tth\,,b\bar b}^{\rm ATLAS}$ (left) and
$\mu_{tth\,,\bar b}^{\rm CMS}$ (right) at $\sqrt{s}=8$ TeV
by stacking the various $thX$ contributions
on the $t\bar t h$ one
for $C_t^S=+1\,,-1\,,+1.5\,,-1.5$ from left to right.
The grey columns in the center without $C_t^S$ value represent  the current
8 TeV LHC data, see Table~\ref{CMS-ATLAS1-table}.

Before closing this section,
we would like to make a comment on the 13 TeV data on 
$\mu_{tth}$.
With $13.3$ fb$^{-1}$ at 13 TeV, ATLAS gives~\cite{atlas-tth-13}:
$$
\mu^{\rm ATLAS}_{tth,multileptons}=2.5^{+1.3}_{-1.1}\,; \ \ \
\mu^{\rm ATLAS}_{tth,\gamma\gamma}=-0.3^{+1.2}_{-1.0}\,; \ \ \
\mu^{\rm ATLAS}_{tth,bb}=2.1^{+1.0}_{-0.9}
$$
leading to the combined value of
$\mu^{\rm ATLAS}_{tth,combined}=1.8^{+0.7}_{-0.7}$.
While, with $12.9$ fb$^{-1}$ at 13 TeV, CMS gives~\cite{cms-tth-13}:
$$
\mu^{\rm CMS}_{tth,multileptons}=2.3^{+0.9}_{-0.8}\,; \ \ \
\mu^{\rm CMS}_{tth,\gamma\gamma(lep)}=1.15^{+2.0}_{-1.4}\,, \ \ \
\mu^{\rm CMS}_{tth,\gamma\gamma(had)}=2.10^{+1.6}_{-1.2}\,; \ \ \
\mu^{\rm CMS}_{tth,bb}=-0.19^{+0.80}_{-0.81}\,.
$$
We observe that
both ATLAS and CMS collaborations again reported
the excesses with a 
significance of about $2\sigma$ in the Higgs decay modes into multileptons.
On the other hand, only CMS (ATLAS) is reporting a significance of about
$2\sigma$ in the $\gamma\gamma$ ($bb$) mode.
Taking a closer look into the $\gamma\gamma$ mode,
we find that our $\gamma\gamma$ results show good 
agreement with
the CMS data, see Table~\ref{sumrr-CMS}.
Though the errors are still large, it is interesting to note that
our results
$\mu_{tth,\gamma\gamma (lep)}^{\rm CMS}=1.05$  and
$\mu_{tth,\gamma\gamma (had)}^{\rm CMS}=2.50$ for $C_t^S=-1$
reproduces the 13-TeV CMS central values.
While, our combined ATLAS results
$1.03,\ 1.37,\ 2.38$ for $C^S_t=(1,\ -1,\ 1.5)$,
see Table~\ref{sumrr-ATLAS},
are in tension with the ATLAS 13 TeV data.
On the other hand, in the $b\bar{b}$ channel,
our results are compatible with the 13 TeV data only in the ATLAS case.

\section{Disentangling $thX$ from $t\bar th$ }

In this section, we show kinematic distributions for the $t\bar{t} h$
and for $thX$ processes in the presence of 
anomalous top-Yukawa coupling in an attempt to 
disentangle $thX$ production from $t\bar{t} h$ one
using specific selection cuts. 
We focus on the $h\rightarrow \gamma\gamma$ channel at
the LHC with $\sqrt{s}=13$ TeV (LHC-13) adopting
the Delphes ATLAS fast detector simulation.
We closely follow the analysis in a previous work \cite{Chang:2014rfa}.
Here we use the $thj$ process for illustration while the other
$thX$ processes have similar features.

\subsection{LHC-13}
In Table~\ref{R(thx13)}, we show the cross sections ratios
$R(t\bar t h)$ and $R(thX)$ with $X=j,jb,W$ at the 13 TeV LHC 
taking $C_t^S = \pm 1$ and $\pm 1.5$. 
Comparing the ratios at $\sqrt{s}=8$ TeV presented in Table~\ref{R(thx)},
we observe the LHC-13 ratios are more or less similar to the
LHC-8 ones.
\begin{table}[t!]
%\begin{table}[th!]
\caption{\small \label{R(thx13)}
The cross-section ratios
$R(t\bar t h)$ and $R(thX)$ with $X=j,jb,W$ defined in Eq.~(\ref{R-thX}).
We are taking $\sqrt{s}=13$ TeV (LHC-13) and
$C_t^S = \pm 1\,, \pm 1.5$.
}
\medskip
\begin{ruledtabular}
\begin{tabular}{lcccc}
 LHC-13 & \multicolumn{4}{c}{With ATLAS Analysis Cuts} \\
\hline
  & $C^{S}_{t}=1$ & $C^{S}_{t}=-1$ & $C^{S}_{t}=1.5$ & $C^{S}_{t}=-1.5$ \\
\hline
Cross Section of $t\bar{t} h$(pb) &0.52&\\%0.52&1.17&1.17\\
$ R(t\bar{t}h) $ & 1&1&2.26&2.26 \\
$ R(thj) $ & 8.31e-2 & 0.97 & 0.14 &  1.51 \\
$ R(thjb) $ & 4.56e-2 & 0.52 & 8.22e-2 & 0.82 \\
$ R(thW) $ & 4.4e-2&0.29&9.39e-2&0.46 \\
\end{tabular}
\end{ruledtabular}
\end{table}

\begin{figure}[th!]
%\begin{figure}[t!]
\centering
\includegraphics[width=1.5in]{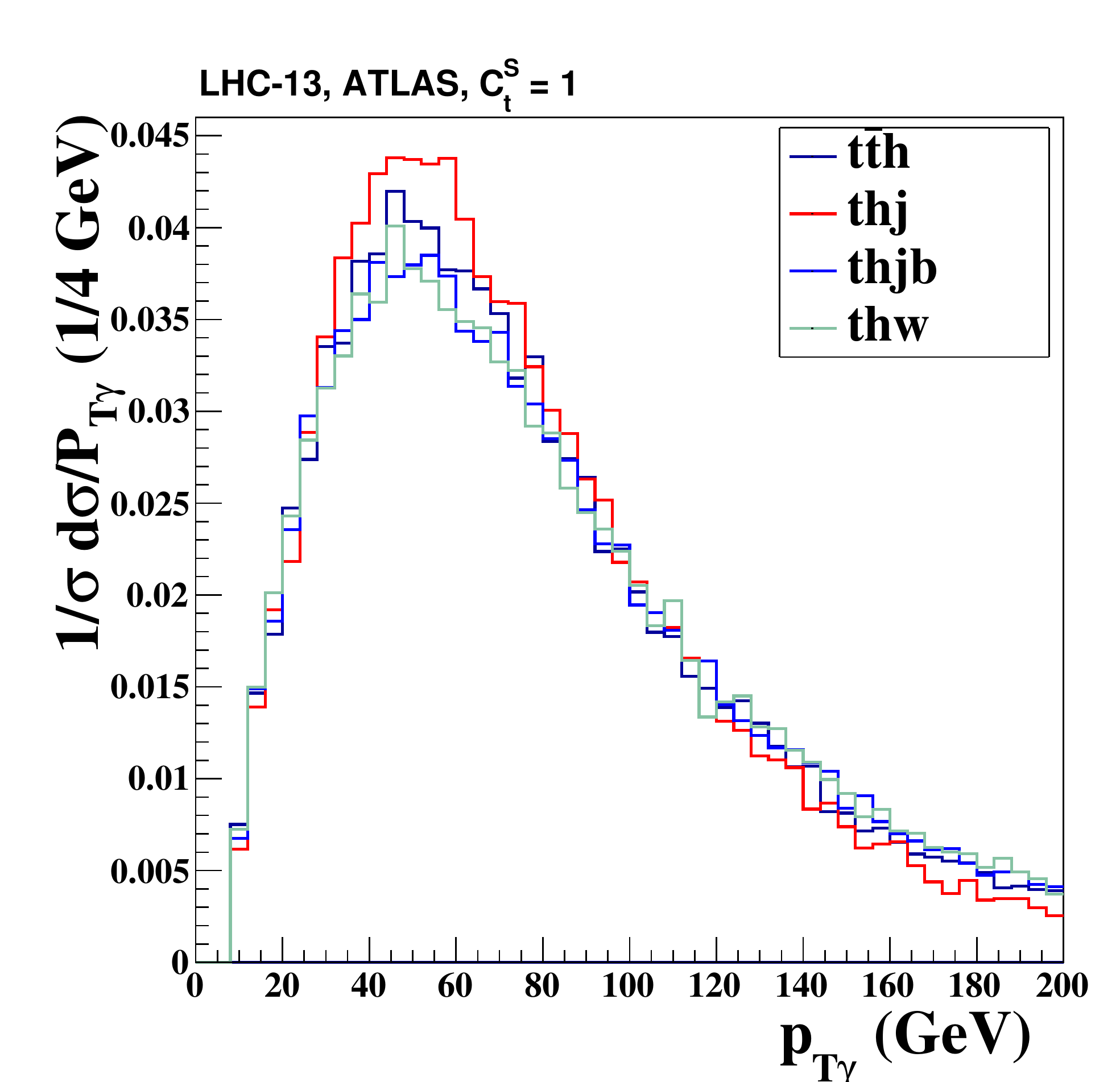}
\includegraphics[width=1.5in]{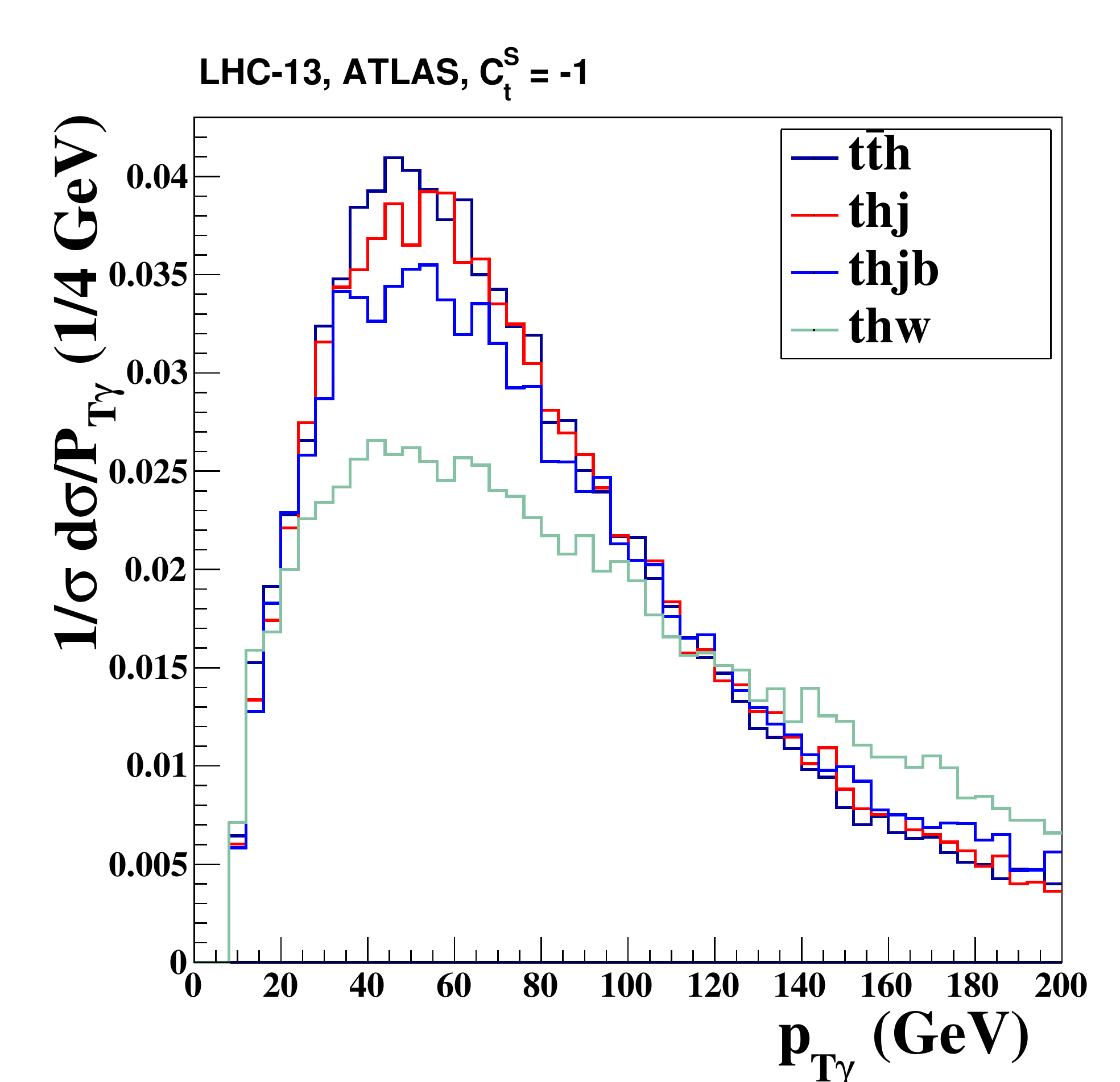}
\includegraphics[width=1.5in]{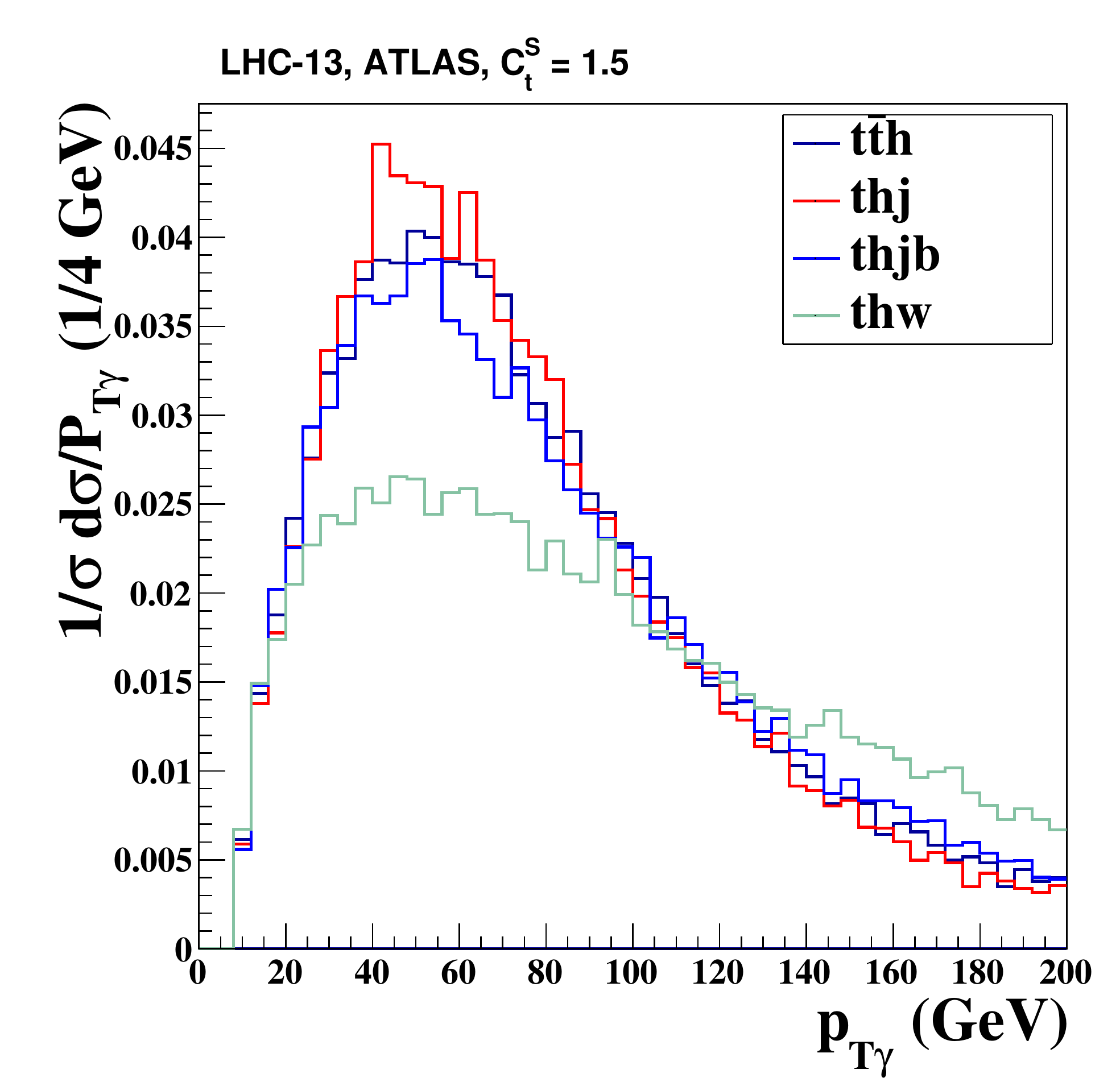}
\includegraphics[width=1.5in]{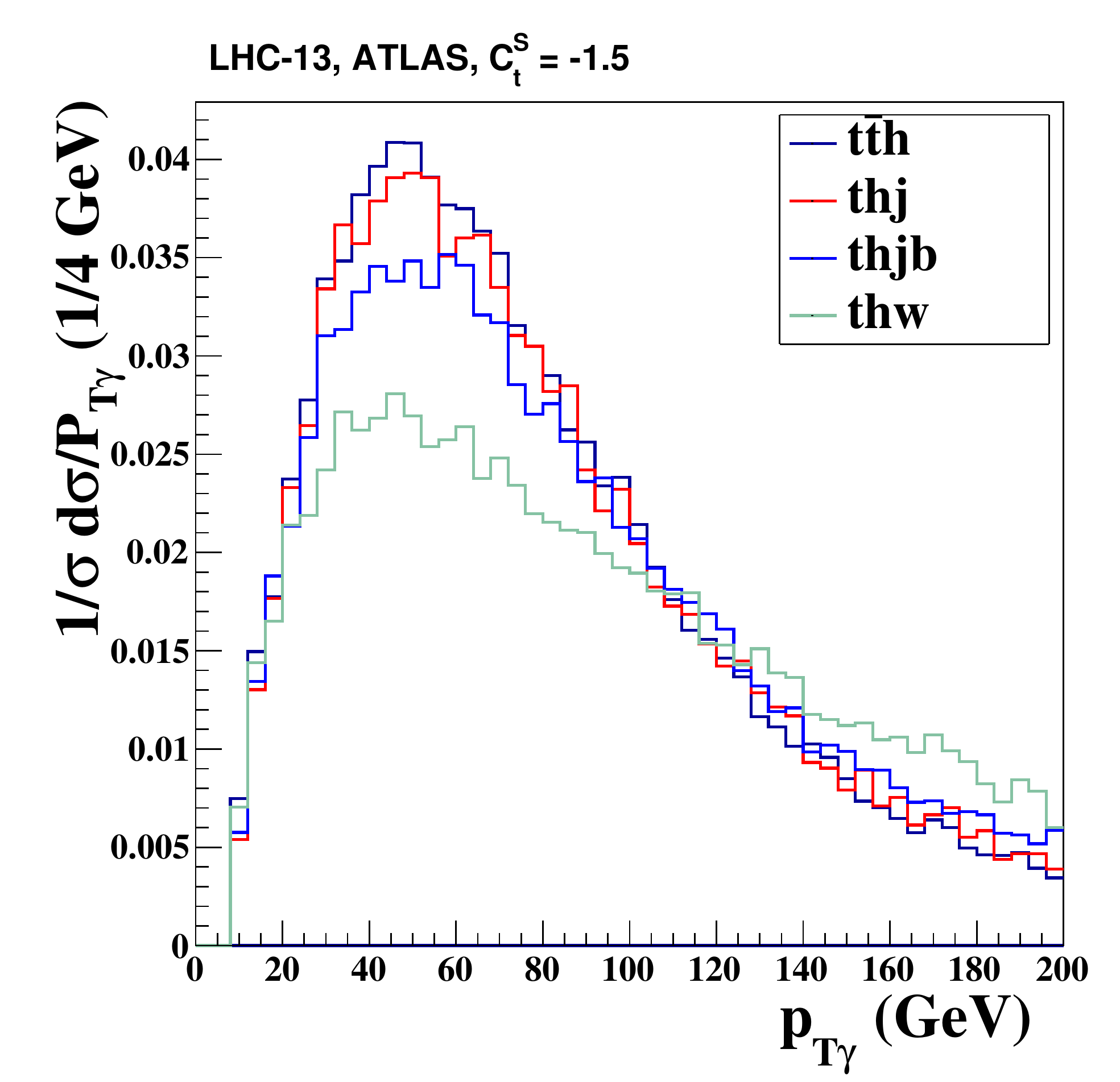}
\caption{\small \label{pt_a_yt}
The $P_{T_\gamma}$ distributions for the $t\bar{t} h$
and $thX$ processes in the $h\to \gamma\gamma$ channel at LHC-13
taking $C_t^S=+1\,,-1\,,+1.5\,,-1.5$ from left to right. 
We use the Delphes ATLAS template for detector simulations.
}\end{figure}

\begin{figure}[th!]
%\begin{figure}[t!]
\centering
\includegraphics[width=1.5in]{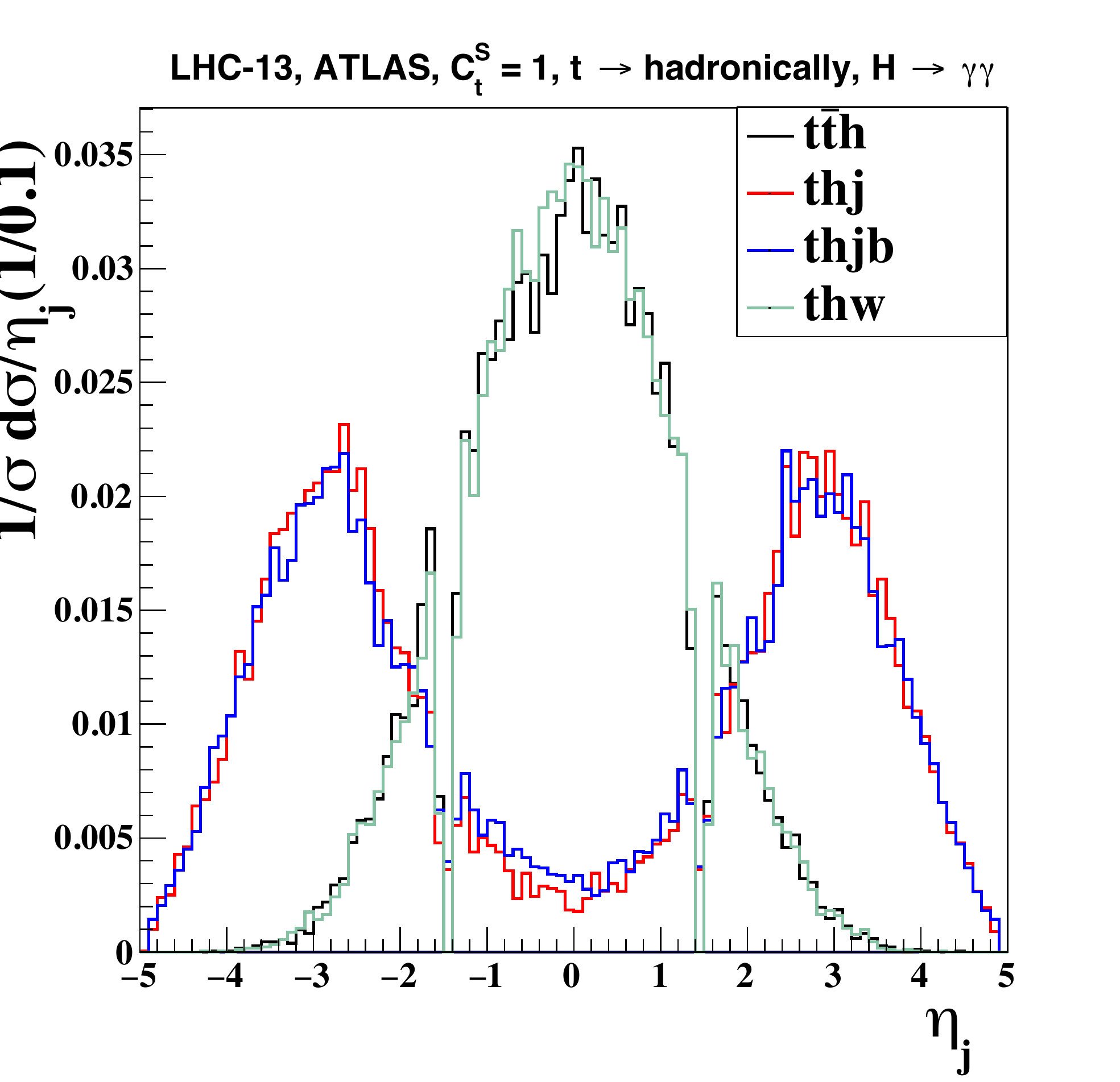}
\includegraphics[width=1.5in]{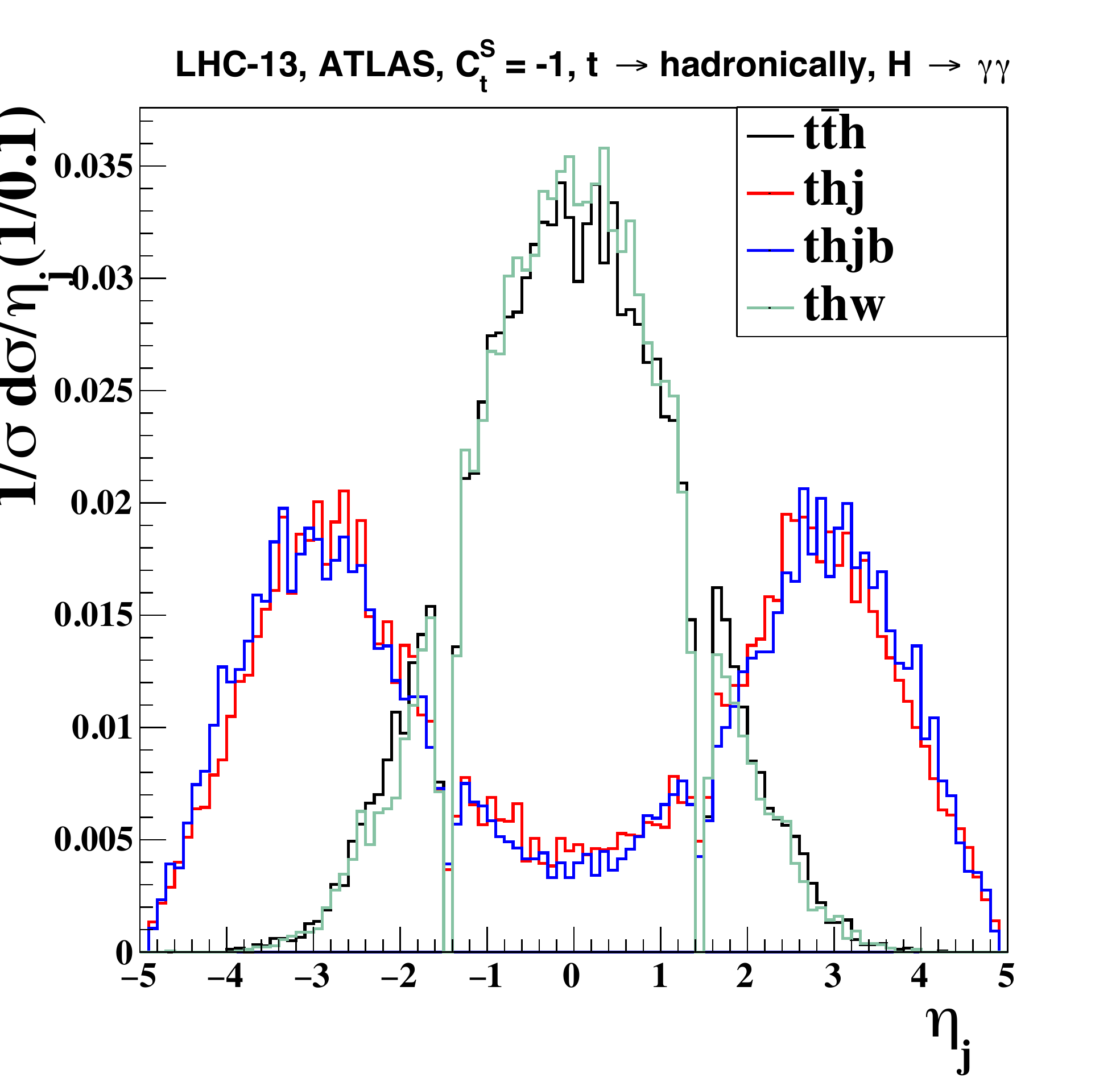}
\includegraphics[width=1.5in]{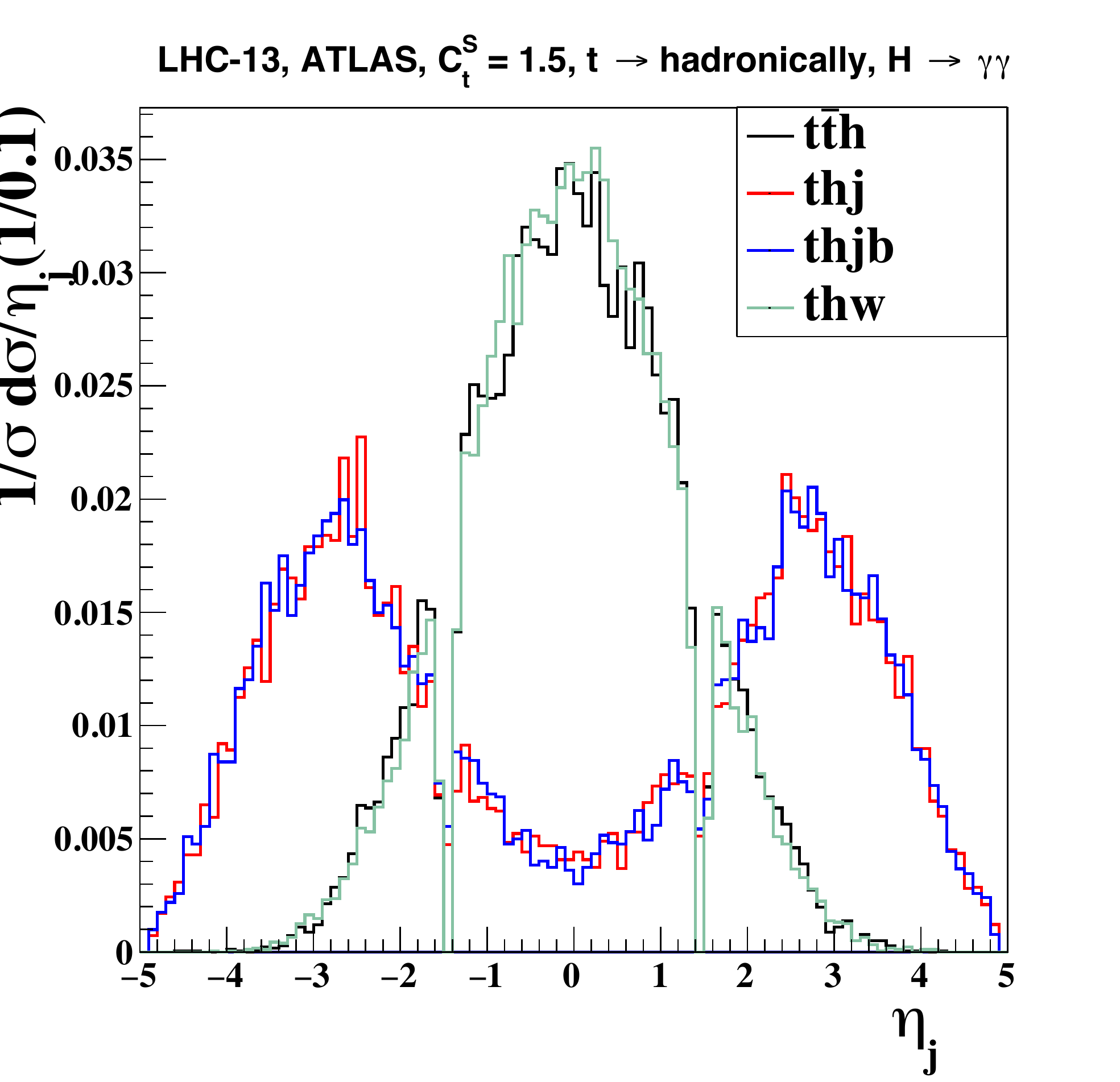}
\includegraphics[width=1.5in]{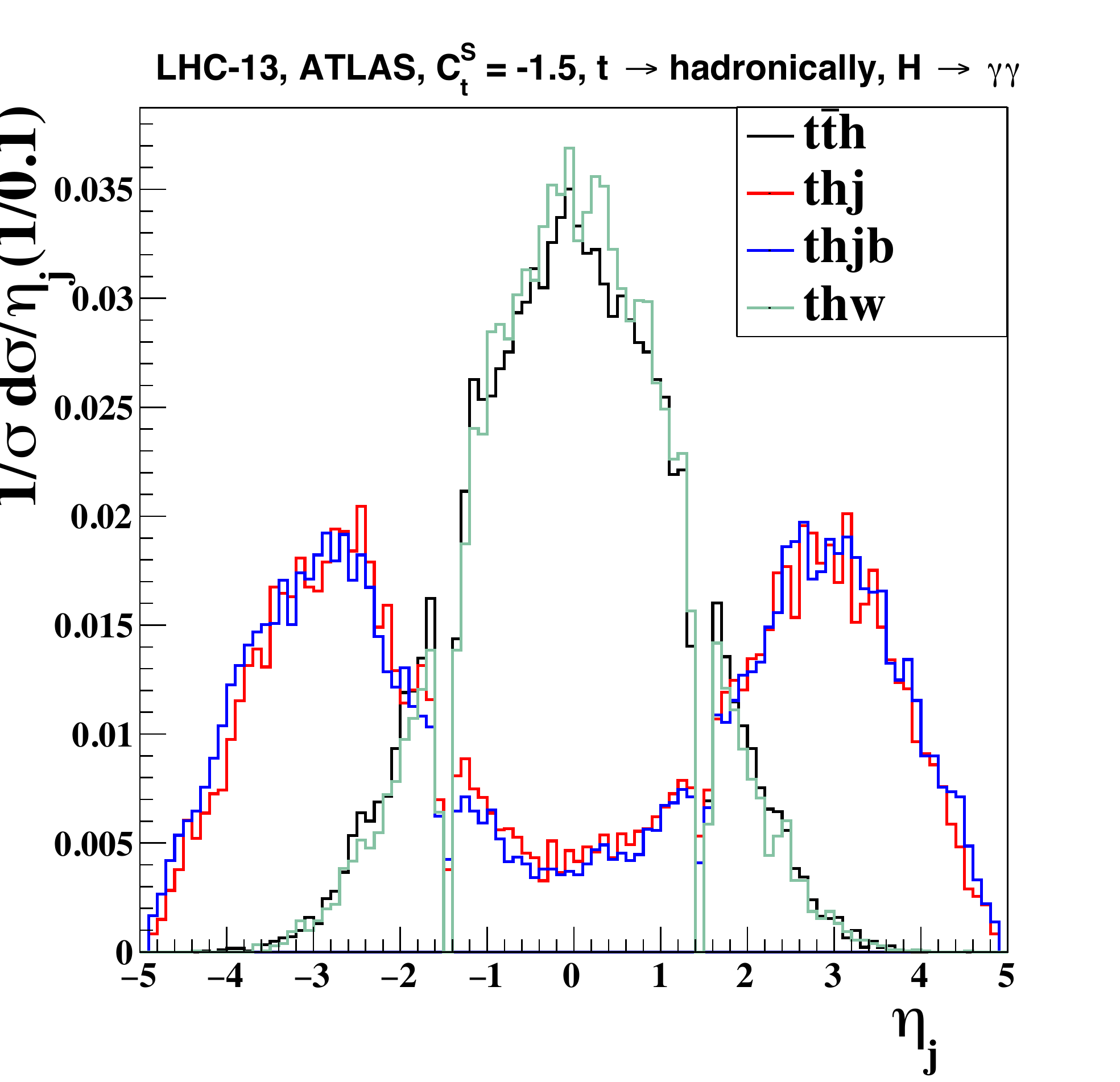}
\caption{\small \label{eta_j_yt}
The same as in Fig.~\ref{pt_a_yt} but for the $\eta_j$ distributions.
%The $\eta_j$ distributions for the $t\bar{t} h$
%and $thX$ processes in the $h\to \gamma\gamma$ channel at LHC-13
%taking $C_t^S=+1\,,-1\,,+1.5\,,-1.5$ from left to right. 
%We use the Delphes ATLAS template for detector simulations.
}\end{figure}

We show the $p_{T_\gamma}$ and $\eta_j$ distributions for the $t\bar th$ and 
$thX$ processes in Fig.~\ref{pt_a_yt} and Fig.~\ref{eta_j_yt}, respectively,
taking $C_t^S= \pm 1\,, \pm 1.5$.
With $C_t^S \neq 1$, the $p_{T_\gamma}$ distribution of the $thX$ process, 
especially, that of the $thW$ process 
becomes harder relative to the $t\bar th$ distribution.
In the $\eta_j$ distributions, the $thj$ and $thjb$ processes
have more forward pseudorapidity.  We therefore come up with a set of
selection cuts summarized in Table~\ref{cut-13},
in which we order the jets according to their energy
since
most of the time the forward jet is the most energetic one. It is 
in general correctly chosen as shown in the $\eta_j$ distribution.
Note that we require to tag one forward jet and apply 
the Higgs-mass window cut on the diphoton invariant mass $M_{\gamma\gamma}$.

%

%\vspace{-1.0cm}
%

%\begin{table}[H]
\begin{table}[t!]
\centering
\caption{\label{cut-13}
Selection cuts to disentangle $thX$ from $t\bar th$.
The Delphes ATLAS template is used.
}
\footnotesize{
  \begin{tabular}{cc}
\hline
LHC-13&{ATLAS Analysis Cuts, $h\rightarrow \gamma\gamma$ } \\
\hline
Basic cuts : &$\Delta R_{ij} > 0.4$ with $i,j$ denoting $b$, $j$ and $l$\\
&$P_{Tb} > 25 GeV, |\eta_b| < 2.5$, $P_{Tl} > 25 GeV, |\eta_l| < 2.5$, $P_{Tj} >
25 GeV, |\eta_j| < 4.7$\\
$\#$ of $\gamma$ $\&$ Higgs mass window cuts :&$N(\gamma) = 2,\
|M_{\gamma\gamma}-m_h| < 5 GeV$\\
$t \bar{t} h$ search &  $|\eta_j|<2.5$ \\
$t h j$ search & Forward jet-tag : $2.5<|\eta_j|<4.7$ \\
\hline
{ $t\rightarrow $ semileptonically $\&$ leptonically} :&$N(e$ or $\mu$)=1,
$E_T^{miss}>20 GeV$, \\
& Invariant Mass cuts for top decay product : $M_{bl} < 200 GeV$\\
&  $t \bar{t} h$ search : $N(j)\ge2,  N(b)\ge 1$ \\
 & $t h j$ search :  $N(j)\le3, N(b)\le 2$  \\

 $t\rightarrow$ hadronically :&$N(e$ or $\mu$)=0, Invariant Mass cuts for top
decay product : $M_{bj_1j_2} < 300 GeV$\\
 &  $t \bar{t} h$ search : $N(j)\ge6, N(b)\ge 2$ \\
 & $t h j$ search : $N(j)\le5, N(b)\le 2$ \\
\hline
  \end{tabular}}
\end{table}

\begin{figure}[t!]
\centering
\includegraphics[width=3in]{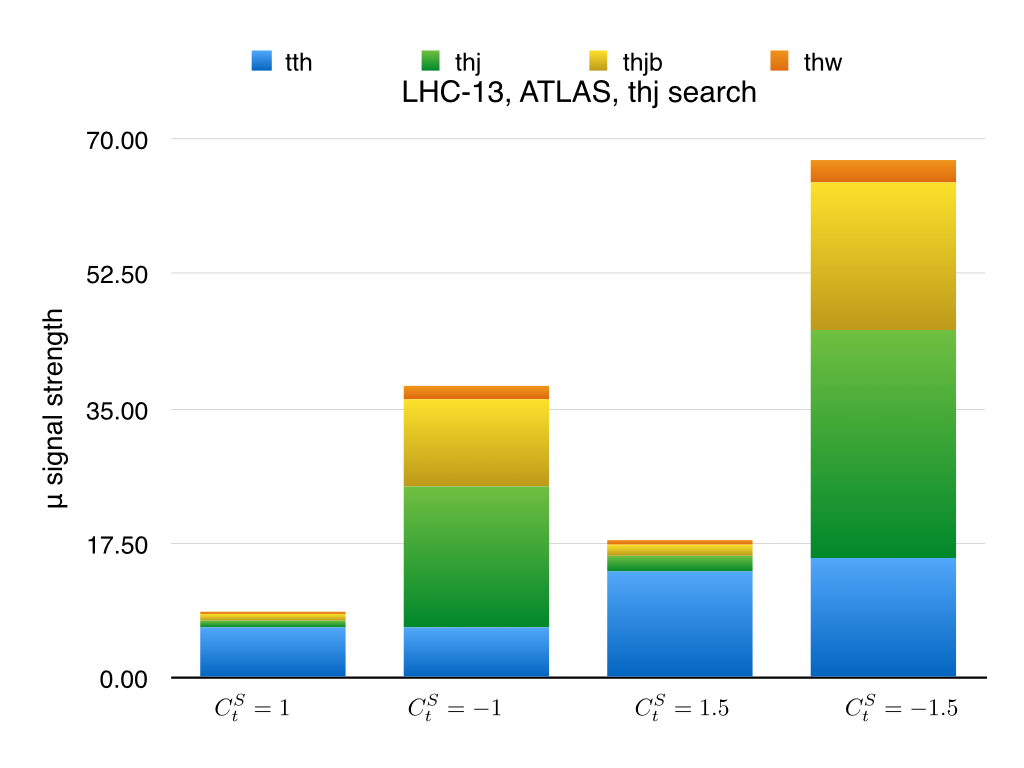}
\includegraphics[width=3in]{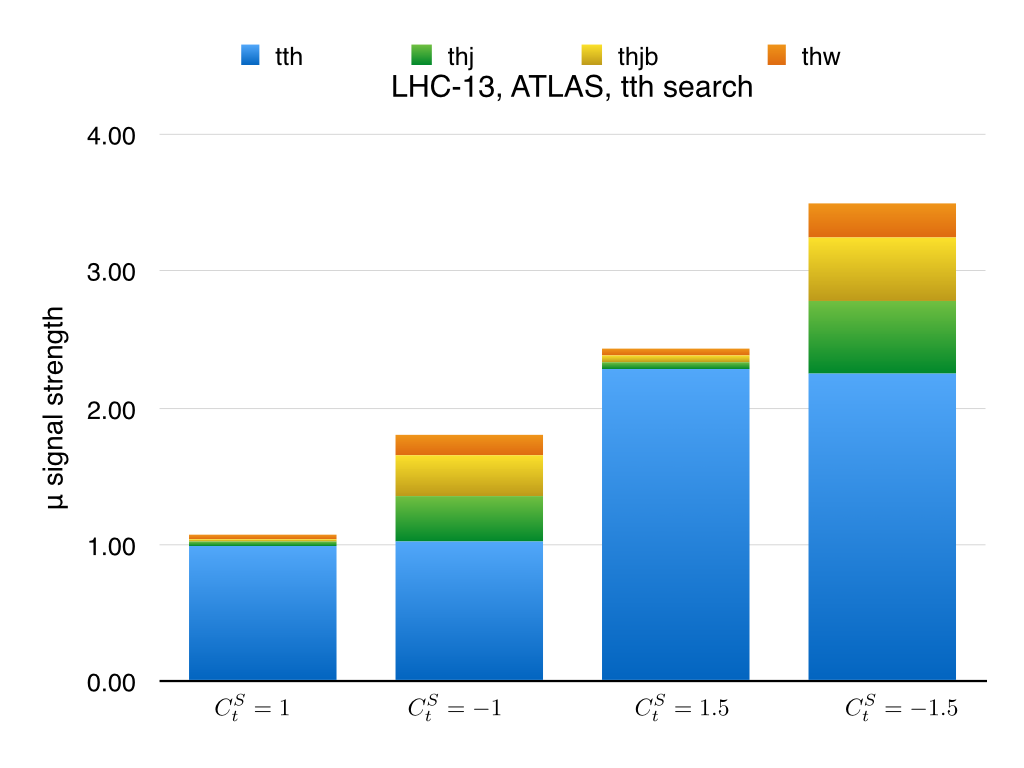}
\caption{\small \label{sum-tth_thj}
Accumulated signal strengths 
$\mu(thj)$ (left) and $\mu(t\bar{t} h)$ (right) at LHC-13 
obtained by stacking the various $thX$ contributions 
on the $t\bar t h$ one
for $C_t^S=+1\,,-1\,,+1.5\,,-1.5$ from left to right.
We use the Delphes ATLAS template for detector simulations.
}\end{figure}

The accumulated $thj$ signal strength $\mu(thj)$
\footnote{
Similarly as $\mu(t\bar{t}h)$ given by Eq.~(\ref{mu-tth}),
the signal strength $\mu(thj)$ is
$$
\mu(thj)= \frac{
\eta_1\sigma(t\bar{t}h)B(t\bar{t}h\to{\cal D})+
\sum_{X=j,jb,W}\eta_X\sigma(thX)B(thX\to{\cal D})}
{\eta_j^{\rm SM}\sigma(thj)_{\rm SM}B(thj\to{\cal D})_{\rm SM}}\,.
$$}
is shown in the left panel of
Fig.~\ref{sum-tth_thj} with $C_t^S=\pm 1\,,\pm 1.5$.
To obtain the signal strength $\mu(thj)$ 
in the $h\to\gamma\gamma$ decay
at LHC-13, we impose the $thj$-specific cuts listed in Table~\ref{cut-13}.
In the right panel of Fig.~\ref{sum-tth_thj}, we show
the accumulated $t\bar th$ signal strength $\mu(t\bar t h)$
obtained by using the $t\bar t h$-specific cuts in the same Table.
We observe that $\mu(thj)$ (left) is dominated by $thj$ (green)
for the negative values of $C_t^S$, implying that our $thj$-specific
cuts are working very efficiently
 when the $thj$ production cross section
is much enhanced with $R(thj)\gsim 1$.
%It is clear from the left panel that using the 
%specific $thj$ cuts the signal strengths are dominated by $thj$ (shown in
%green in the colored bar) for $C_t^S=-1,-1.5$ while dominated by
%$t\bar th$ for $C_t^S = 1,1.5$ (shown in blue). 
%
On the other hand, $\mu(t\bar t h)$ (right) is dominated 
by $t\bar t h$ (blue) independently of $C_t^S$ and we observe that 
our  $t\bar t h$-specific cuts are working reasonably well
as in the LHC-8 case (the left panel of Fig.~\ref{sum-rr-atlas-cms}).
%On the other hand, from the right panel using the specific 
%$t\bar th$ cuts the signal strengths are  dominated by $t\bar th$ 
%for all $C_t^S$ (shown in blue). 
%
We can further draw a few observations from Fig.~\ref{sum-tth_thj} as follows.
\begin{enumerate}

\item
When the experiment is targeting at 
$t\bar th$ production using 
 the $t\bar t h$-specific cuts, 
there are contaminations from the $thX$ processes. 
For positive $C_t^S$, the $thX$ contaminations are small.
But, for negative $C_t^S$, they can be as large as the $t\bar th$ signals.
For $C_t^S = -1$, for example, $\mu(t\bar th) \sim 2$ 
and only half of which
comes from $t\bar th$.

\item From the left panel, we can see  that the $thX$ processes 
dominate the signal strength $\mu(thj)$ for negative $C_t^S$, 
which means that 
the $thj$-specific selection cuts we employed
indeed can single out the $thj$ process from the $t\bar th$ one. 

\item 
The large values of $\mu(thj) \sim {\cal O}(10)$ when $C_t^S$ deviates from
its SM value $1$ imply 
that the direct $thj$ searches are also 
important as complementary channels.
Current LHC constraints on the $thj$ searches at $\sqrt{s}=8$ TeV in 
are still weak~\cite{thj-8}, 
so that more data are needed at 
$\sqrt{s}=13$ TeV in the future to 
probe the anomalous top-Yukawa coupling
through this channel.

\end{enumerate}

%\begin{figure}[th!]

\section{Conclusions}
Usually, the associated Higgs production with a single top quark 
dubbed as $thX$ with $X=j,jb,W$ makes 
only small contributions to the overall experimental 
signal strength of $\mu(t\bar th)$.
In this work, however, we have demonstrated explicitly that 
the $thX$ processes can significantly increase the experimentally measured
signal strength $\mu(t\bar th)$ 
when the relative sign of the top-Yukawa coupling 
to the gauge-Higgs coupling is reversed.
Furthermore, we have shown explicitly that
the $thX$ processes  contaminate
at quite different levels in various detection modes of $tth$,
depending on the value of top-Yukawa coupling, on the cuts
used in each experiment, and on the decay mode of the Higgs
boson.  Such behavior is
far more complicated than simply assuming a small constant
level of contamination in all channels.
The signal strengths can be as large as
$2 - 4$ in the category Leptons for $h\to$ multileptons, 
$2 - 4.5$
in the category $\gamma\gamma$ for $h\to\gamma\gamma$, and 
$2 - 4$ in the category $b\bar b$ for  $h\to b\bar b$. 
Assuming the mild excesses observed in $t\bar t h$ production
at the LHC are real,
we note that all go in the right direction to match them.
%We note that all go in the right direction to match the current 
%8-TeV data on the mild excesses observed at the LHC.

When more data are collected at $\sqrt{s}=13$ TeV, we can choose
more specific cuts to single out the $thX$ processes, which can 
effectively determine the size and the sign of the top-Yukawa coupling.

We offer the following comments on our findings.

\begin{enumerate}

\item
The current data on $t\bar th$ production showed mild excesses at some level
\footnote{
In the 13 TeV data, only the multilepton channel shows
the mild excess both in ATLAS and CMS.
On the other hand, a similar excess in the $bb$ ($\gamma\gamma$) channel
is reported only by ATLAS (CMS).}.
Although they may be simply due to statistical fluctuations,
in this work,
we have taken the liberty of interpreting the mild excesses by 
exploiting the strong entanglement between $thX$ and $t\bar t h$.
Our case study
would be very useful if the future data support the excesses.

\item When the top-Yukawa coupling is kept at the SM value, i.e.
$C_t^S = 1$, the contamination from all the $thX$ processes is small,
only about $5-15\%$, and that can be regarded as 
a sort of small higher-order corrections.

\item However, when the sign of the top-Yukawa coupling is reversed, i.e.
$C_t^S = -1$, the $thX$ contributions are significantly enhanced.  
And the resulting signal strengths can be as large as 
$1.4-2.0$ (category Leptons),
$1.0-2.5$ (category $\gamma\gamma$), and
$1.4-2.0$ (category $b\bar b$),
explaining the experimental excesses shown 
in Table~\ref{CMS-ATLAS1-table}.

\item When $C_t^S$ is further negative, say $-1.5$, the resulting
signal strength $\mu(t\bar th)$ further increases to 
$2.7-3.9$ (category Leptons),
$2.4-4.5$ (category $\gamma\gamma$), and
$2.6-3.8$ (category $b\bar b$).

\item 
In the approach adopted in this work,
the dominant $thX$ processes are $thj$ and $thjb$ both of which
contain a very forward energetic jet. 
Also, as shown in Fig.~\ref{pt_a_yt},
the $thW$ process has a harder $p_T$ photon.  
Therefore, we successfully
come up with a set of selection cuts to single out the $thX$ processes
from the $t\bar t h$ process. It has been shown clearly in the left panel
of Fig.~\ref{sum-tth_thj}.

\item
One very useful observation in our work is that 
the contributions from various production processes of $t\bar th$, $thj$, 
$thjb$, and $thW$ to the accumulated signal strengths 
strongly depend not only on the Higgs decay channels but also on
the experiment (ATLAS or CMS),
as can be seen in Figs.~\ref{sum-1-atlas-cms} -- \ref{sum-bb-atlas-cms}.

\end{enumerate}

\section*{Acknowledgment}  

K.C. was supported by the MoST of Taiwan under Grants number 
102-2112-M-007-015-MY3.
J.S.L. was supported by
the National Research Foundation of Korea (NRF) grant
No. NRF-2016R1E1A1A01943297.
%
%J.S.L thanks National Center for Theoretical Sciences (Hsinchu, Taiwan) for
%the great hospitality extended to him while this work was being
%performed.

%%%%%%%%%%%%%%%%%%%%%%%%%%%%%%%%%%%%%%%%


\begin{thebibliography}{99}

\bibitem{atlas}
G.~Aad {\it et al.}  [ATLAS Collaboration],
  %``Observation of a new particle in the search for the Standard Model Higgs boson with the ATLAS detector at the LHC,''
  Phys.\ Lett.\ B {\bf 716}, 1 (2012)
  [arXiv:1207.7214 [hep-ex]].
  %%CITATION = ARXIV:1207.7214;%%

\bibitem{cms}
S.~Chatrchyan {\it et al.}  [CMS Collaboration],
  %``Observation of a new boson at a mass of 125 GeV with the CMS experiment at the LHC,''
  Phys.\ Lett.\ B {\bf 716}, 30 (2012)
  [arXiv:1207.7235 [hep-ex]].
  %%CITATION = ARXIV:1207.7235;%%

\bibitem{higgcision}See for example,
  K.~Cheung, J.~S.~Lee and P.~-Y.~Tseng,
  %``Higgs Precision (Higgcision) Era begins,''
  JHEP {\bf 1305}, 134 (2013)
  [arXiv:1302.3794 [hep-ph]].
  %%CITATION = ARXIV:1302.3794;%%

  K.~Cheung, J.~S.~Lee and P.~Y.~Tseng,
  %``Higgs precision analysis updates 2014,''
  Phys.\ Rev.\ D {\bf 90}, 095009 (2014)
  doi:10.1103/PhysRevD.90.095009
  [arXiv:1407.8236 [hep-ph]].

\bibitem{higgs}
P.~W.~Higgs,
  %``Broken Symmetries and the Masses of Gauge Bosons,''
  Phys.\ Rev.\ Lett.\  {\bf 13}, 508 (1964);
  %%CITATION = PRLTA,13,508;%%
F.~Englert and R.~Brout,
  %``Broken Symmetry and the Mass of Gauge Vector Mesons,''
  Phys.\ Rev.\ Lett.\  {\bf 13}, 321 (1964);
  %%CITATION = PRLTA,13,321;%%
G.~S.~Guralnik, C.~R.~Hagen and T.~W.~B.~Kibble,
  %``Global Conservation Laws and Massless Particles,''
  Phys.\ Rev.\ Lett.\  {\bf 13}, 585 (1964).
  %%CITATION = PRLTA,13,585;%%

\bibitem{Khachatryan:2014jba}
ATLAS Coll., "Measurements of the Higgs boson production and decay rates and couplings using pp collision data at sqrt(s)=7 and 8 TeV in the ATLAS experiment", ATLAS-CONF-2015-007 (March 2015);
V.~Khachatryan {\it et al.} [CMS Collaboration],
  %``Precise determination of the mass of the Higgs boson and tests of compatibility of its couplings with the standard model predictions using proton collisions at 7 and 8 $\,\text {TeV}$,''
  Eur.\ Phys.\ J.\ C {\bf 75}, no. 5, 212 (2015)
  doi:10.1140/epjc/s10052-015-3351-7
  [arXiv:1412.8662 [hep-ex]].

\bibitem{Aad:2015gdg} 
  G.~Aad {\it et al.} [ATLAS Collaboration],
  %``Analysis of events with $b$-jets and a pair of leptons of the same charge in $pp$ collisions at $\sqrt{s}=8$ TeV with the ATLAS detector,''
  JHEP {\bf 1510}, 150 (2015)
  doi:10.1007/JHEP10(2015)150
  [arXiv:1504.04605 [hep-ex]].
  
\bibitem{Khachatryan:2014qaa} 
  V.~Khachatryan {\it et al.} [CMS Collaboration],
  %``Search for the associated production of the Higgs boson with a top-quark pair,''
  JHEP {\bf 1409}, 087 (2014)
  [JHEP {\bf 1410}, 106 (2014)]
  doi:10.1007/JHEP09(2014)087, 10.1007/JHEP10(2014)106
  [arXiv:1408.1682 [hep-ex]].  
  
\bibitem{ss2l-ex}
  S.~Chatrchyan {\it et al.} [CMS Collaboration],
  %``Search for new physics in events with same-sign dileptons and jets in pp collisions at $\sqrt{s}$ = 8 TeV,''
  JHEP {\bf 1401}, 163 (2014)
  [JHEP {\bf 1501}, 014 (2015)]
  doi:10.1007/JHEP01(2015)014, 10.1007/JHEP01(2014)163
  [arXiv:1311.6736, arXiv:1311.6736 [hep-ex]].

  G.~Aad {\it et al.} [ATLAS Collaboration],
  %``Search for supersymmetry at $\sqrt{s}$=8 TeV in final states with jets and two same-sign leptons or three leptons with the ATLAS detector,''
  JHEP {\bf 1406}, 035 (2014)
  doi:10.1007/JHEP06(2014)035
  [arXiv:1404.2500 [hep-ex]].

  G.~Aad {\it et al.} [ATLAS Collaboration],
  %``Search for the associated production of the Higgs boson with a top quark pair in multilepton final states with the ATLAS detector,''
  Phys.\ Lett.\ B {\bf 749}, 519 (2015)
  doi:10.1016/j.physletb.2015.07.079
  [arXiv:1506.05988 [hep-ex]].

  G.~Aad {\it et al.} [ATLAS Collaboration],
  %``Measurement of the $ t\bar{t}W $ and $ t\bar{t}Z $ production cross sections in pp collisions at $ \sqrt{s}=8 $ TeV with the ATLAS detector,''
  JHEP {\bf 1511}, 172 (2015)
  doi:10.1007/JHEP11(2015)172
  [arXiv:1509.05276 [hep-ex]].

  V.~Khachatryan {\it et al.} [CMS Collaboration],
  %``Observation of top quark pairs produced in association with a vector boson in pp collisions at sqrt(s) = 8 TeV,''
  arXiv:1510.01131 [hep-ex].

\bibitem{ss2l-ph}
  B.~Bhattacherjee, S.~Chakraborty and S.~Mukherjee,
  %``$H \rightarrow \tau \mu$ and excess in $t\bar{t}h$: Connecting the dots in the hope for the first glimpse of BSM Higgs signal,''
  arXiv:1505.02688 [hep-ph].
  
  P.~Huang, A.~Ismail, I.~Low and C.~E.~M.~Wagner,
  %``Same-Sign Dilepton Excesses and Light Top Squarks,''
  Phys.\ Rev.\ D {\bf 92}, no. 7, 075035 (2015)
  doi:10.1103/PhysRevD.92.075035
  [arXiv:1507.01601 [hep-ph]].

  C.~R.~Chen, H.~C.~Cheng and I.~Low,
  %``Same-Sign Dilepton Excesses and Vector-like Quarks,''
  arXiv:1511.01452 [hep-ph].

\bibitem{Aad:2015iha} 
  G.~Aad {\it et al.} [ATLAS Collaboration],
  %``Search for the associated production of the Higgs boson with a top quark pair in multilepton final states with the ATLAS detector,''
  Phys.\ Lett.\ B {\bf 749}, 519 (2015)
  doi:10.1016/j.physletb.2015.07.079
  [arXiv:1506.05988 [hep-ex]].

\bibitem{Chang:2014rfa} 
  J.~Chang, K.~Cheung, J.~S.~Lee and C.~T.~Lu,
  %``Probing the Top-Yukawa Coupling in Associated Higgs production with a Single Top Quark,''
  JHEP {\bf 1405}, 062 (2014)
  doi:10.1007/JHEP05(2014)062
  [arXiv:1403.2053 [hep-ph]].

\bibitem{singletop.old}  
  T.~M.~P.~Tait and C.-P.~Yuan,
  %``Single top quark production as a window to physics beyond the standard model,''
  Phys.\ Rev.\ D {\bf 63}, 014018 (2000)
  doi:10.1103/PhysRevD.63.014018
  [hep-ph/0007298].

%\cite{Stirling:1992fx}
%\bibitem{Stirling:1992fx}
  W.~J.~Stirling and D.~J.~Summers,
  %``Production of an intermediate mass Higgs boson in association with a
  %single top quark at LHC and SSC,''
  Phys.\ Lett.\ B {\bf 283} (1992) 411.
  doi:10.1016/0370-2693(92)90040-B
  %%CITATION = doi:10.1016/0370-2693(92)90040-B;%%
  %29 citations counted in INSPIRE as of 11 Oct 2016

%\cite{Ballestrero:1992bk}
%\bibitem{Ballestrero:1992bk}
  A.~Ballestrero and E.~Maina,
  %``t anti-b H production for an intermediate mass Higgs,''
  Phys.\ Lett.\ B {\bf 299} (1993) 312.
  doi:10.1016/0370-2693(93)90265-J
  %%CITATION = doi:10.1016/0370-2693(93)90265-J;%%
  %25 citations counted in INSPIRE as of 11 Oct 2016

%\cite{Bordes:1992jy}
%\bibitem{Bordes:1992jy}
  G.~Bordes and B.~van Eijk,
  %``On the associate production of a neutral intermediate mass Higgs boson
  %with a single top quark at the LHC and SSC,''
  Phys.\ Lett.\ B {\bf 299} (1993) 315.
  doi:10.1016/0370-2693(93)90266-K
  %%CITATION = doi:10.1016/0370-2693(93)90266-K;%%
  %27 citations counted in INSPIRE as of 11 Oct 2016

%\cite{Maltoni:2001hu}
%\bibitem{Maltoni:2001hu}
  F.~Maltoni, K.~Paul, T.~Stelzer and S.~Willenbrock,
  %``Associated production of Higgs and single top at hadron colliders,''
  Phys.\ Rev.\ D {\bf 64} (2001) 094023
  doi:10.1103/PhysRevD.64.094023
  [hep-ph/0106293].
  %%CITATION = doi:10.1103/PhysRevD.64.094023;%%
  %73 citations counted in INSPIRE as of 11 Oct 2016

 V.~Barger, M.~McCaskey and G.~Shaughnessy,
  %``Single top and Higgs associated production at the LHC,''
  Phys.\ Rev.\ D {\bf 81}, 034020 (2010)
  [arXiv:0911.1556 [hep-ph]].
  %%CITATION = ARXIV:0911.1556;%%
  
\bibitem{th-others}  
  S.~Biswas, E.~Gabrielli and B.~Mele,
  %``Single top and Higgs associated production as a probe of the Htt coupling sign at the LHC,''
  JHEP {\bf 1301}, 088 (2013)
  [arXiv:1211.0499 [hep-ph]]; 
  %%CITATION = ARXIV:1211.0499;%%
  S.~Biswas, E.~Gabrielli, F.~Margaroli and B.~Mele,
  %``Direct constraints on the top-Higgs coupling from the 8 TeV LHC data,''
  JHEP {\bf 07}, 073 (2013)
  [arXiv:1304.1822 [hep-ph]].
  %%CITATION = ARXIV:1304.1822;%%

  M.~Farina, C.~Grojean, F.~Maltoni, E.~Salvioni and A.~Thamm,
  %``Lifting degeneracies in Higgs couplings using single top production in association with a Higgs boson,''
  JHEP {\bf 1305}, 022 (2013)
  [arXiv:1211.3736 [hep-ph]].
  %%CITATION = ARXIV:1211.3736;%%

  P.~Agrawal, S.~Mitra and A.~Shivaji,
  %``Effect of Anomalous Couplings on the Associated Production of a Single Top Quark and a Higgs Boson at the LHC,''
  arXiv:1211.4362 [hep-ph].
  %%CITATION = ARXIV:1211.4362;%%

  J.~Ellis, D.~S.~Hwang, K.~Sakurai and M.~Takeuchi,
  %``Disentangling Higgs-Top Couplings in Associated Production,''
  arXiv:1312.5736 [hep-ph].
  %%CITATION = ARXIV:1312.5736;%%
  
  C.~Englert and E.~Re,
  %``Bounding the top Yukawa with Higgs-associated single-top production,''
  arXiv:1402.0445 [hep-ph].
  %%CITATION = ARXIV:1402.0445;%%
  %1 citations counted in INSPIRE as of 05 Mar 2014

  A.~Kobakhidze, L.~Wu and J.~Yue,
  %``Anomalous Top-Higgs Couplings and Top Polarisation in Single Top and Higgs Associated Production at the LHC,''
  JHEP {\bf 1410}, 100 (2014)
  doi:10.1007/JHEP10(2014)100
  [arXiv:1406.1961 [hep-ph]].

%\cite{Demartin:2014fia}
%\bibitem{Demartin:2014fia}
  F.~Demartin, F.~Maltoni, K.~Mawatari, B.~Page and M.~Zaro,
  %``Higgs characterisation at NLO in QCD: CP properties of the top-quark Yukawa interaction,''
  Eur.\ Phys.\ J.\ C {\bf 74} (2014) no.9,  3065
  doi:10.1140/epjc/s10052-014-3065-2
  [arXiv:1407.5089 [hep-ph]].
  %%CITATION = doi:10.1140/epjc/s10052-014-3065-2;%%
  %40 citations counted in INSPIRE as of 26 Jul 2016

%\cite{Khatibi:2014bsa}
%\bibitem{Khatibi:2014bsa}
  S.~Khatibi and M.~Mohammadi Najafabadi,
  %``Exploring the Anomalous Higgs-top Couplings,''
  Phys.\ Rev.\ D {\bf 90} (2014) no.7,  074014
  doi:10.1103/PhysRevD.90.074014
  [arXiv:1409.6553 [hep-ph]].
  %%CITATION = doi:10.1103/PhysRevD.90.074014;%%
  %19 citations counted in INSPIRE as of 26 Jul 2016

  J.~Yue,
  %``Enhanced $thj$ signal at the LHC with $h\rightarrow \gamma\gamma$ decay and $\mathcal{CP}$-violating top-Higgs coupling,''
  Phys.\ Lett.\ B {\bf 744}, 131 (2015)
  doi:10.1016/j.physletb.2015.03.044
  [arXiv:1410.2701 [hep-ph]].

  F.~Demartin, F.~Maltoni, K.~Mawatari and M.~Zaro,
  %``Higgs production in association with a single top quark at the LHC,''
  Eur.\ Phys.\ J.\ C {\bf 75}, no. 6, 267 (2015)
  doi:10.1140/epjc/s10052-015-3475-9
  [arXiv:1504.00611 [hep-ph]].

%\cite{Buckley:2015vsa}
%\bibitem{Buckley:2015vsa}
  M.~R.~Buckley and D.~Goncalves,
  %``Boosting the Direct CP Measurement of the Higgs-Top Coupling,''
  Phys.\ Rev.\ Lett.\  {\bf 116} (2016) no.9,  091801
  doi:10.1103/PhysRevLett.116.091801
  [arXiv:1507.07926 [hep-ph]].
  %%CITATION = doi:10.1103/PhysRevLett.116.091801;%%
  %22 citations counted in INSPIRE as of 26 Jul 2016

%\cite{Goncalves:2015mfa}
%\bibitem{Goncalves:2015mfa}
  D.~Goncalves, F.~Krauss, S.~Kuttimalai and P.~Maierhöfer,
  %``Higgs-Strahlung: Merging the NLO Drell-Yan and Loop-Induced 0+1 jet Multiplicities,''
  Phys.\ Rev.\ D {\bf 92} (2015) no.7,  073006
  doi:10.1103/PhysRevD.92.073006
  [arXiv:1509.01597 [hep-ph]].
  %%CITATION = doi:10.1103/PhysRevD.92.073006;%%
  %9 citations counted in INSPIRE as of 26 Jul 2016

%\cite{Demartin:2016axk}
%\bibitem{Demartin:2016axk}
  F.~Demartin, B.~Maier, F.~Maltoni, K.~Mawatari and M.~Zaro,
  %``tWH associated production at the LHC,''
  arXiv:1607.05862 [hep-ph].
  %%CITATION = ARXIV:1607.05862;%%

%\cite{Rindani:2016scj}
%\bibitem{Rindani:2016scj}
  S.~D.~Rindani, P.~Sharma and A.~Shivaji,
  %``Unraveling the CP phase of top-Higgs coupling in associated production at the LHC,''
  arXiv:1605.03806 [hep-ph].
  %%CITATION = ARXIV:1605.03806;%%
  %1 citations counted in INSPIRE as of 26 Jul 2016

\bibitem{tth:multilepton.old}
%\cite{Maltoni:2002jr}
%\bibitem{Maltoni:2002jr}
  F.~Maltoni, D.~L.~Rainwater and S.~Willenbrock,
  %``Measuring the top quark Yukawa coupling at hadron colliders via
  %$t\bar{t}H,H\to W^+W^-$,''
  Phys.\ Rev.\ D {\bf 66} (2002) 034022
  doi:10.1103/PhysRevD.66.034022
  [hep-ph/0202205];
  %%CITATION = doi:10.1103/PhysRevD.66.034022;%%
  %78 citations counted in INSPIRE as of 11 Oct 2016
%
%\cite{Belyaev:2002ua}
%\bibitem{Belyaev:2002ua}
  A.~Belyaev and L.~Reina,
  %``pp ---> t anti-t H, H ---> tau+ tau-: Toward a model independent
  %determination of the Higgs boson couplings at the LHC,''
  JHEP {\bf 0208} (2002) 041
  doi:10.1088/1126-6708/2002/08/041
  [hep-ph/0205270].
  %%CITATION = doi:10.1088/1126-6708/2002/08/041;%%
  %113 citations counted in INSPIRE as of 11 Oct 2016


%\cite{Lee:2003nta}
\bibitem{Lee:2003nta}
  J.~S.~Lee, A.~Pilaftsis, M.~S.~Carena, S.~Y.~Choi, M.~Drees, J.~R.~Ellis and
C.~E.~M.~Wagner,
%  ``CPsuperH: A Computational tool for Higgs phenomenology in the minimal
%supersymmetric standard model with explicit CP violation,''
  Comput.\ Phys.\ Commun.\  {\bf 156} (2004) 283
  [hep-ph/0307377].
  %%CITATION = HEP-PH/0307377;%%

\bibitem{Aad:2014lma} 
  G.~Aad {\it et al.} [ATLAS Collaboration],
  %``Search for $H \to \gamma\gamma$ produced in association with top quarks and constraints on the Yukawa coupling between the top quark and the Higgs boson using data taken at 7 TeV and 8 TeV with the ATLAS detector,''
  Phys.\ Lett.\ B {\bf 740}, 222 (2015)
  doi:10.1016/j.physletb.2014.11.049
  [arXiv:1409.3122 [hep-ex]].

\bibitem{Aad:2015gra} 
  G.~Aad {\it et al.} [ATLAS Collaboration],
  %``Search for the Standard Model Higgs boson produced in association with top quarks and decaying into $b\bar{b}$ in pp collisions at $\sqrt{s}$ = 8 TeV with the ATLAS detector,''
  Eur.\ Phys.\ J.\ C {\bf 75}, no. 7, 349 (2015)
  doi:10.1140/epjc/s10052-015-3543-1
  [arXiv:1503.05066 [hep-ex]].

\bibitem{Alwall:2014hca} 
  J.~Alwall {\it et al.},
  %``The automated computation of tree-level and next-to-leading order differential cross sections, and their matching to parton shower simulations,''
  JHEP {\bf 1407}, 079 (2014)
  doi:10.1007/JHEP07(2014)079
  [arXiv:1405.0301 [hep-ph]].

\bibitem{Sjostrand:2007gs} 
  T.~Sjostrand, S.~Mrenna and P.~Z.~Skands,
  %``A Brief Introduction to PYTHIA 8.1,''
  Comput.\ Phys.\ Commun.\  {\bf 178}, 852 (2008)
  doi:10.1016/j.cpc.2008.01.036
  [arXiv:0710.3820 [hep-ph]].
  
\bibitem{delphes3}
J.~de Favereau {\it et al.}  [DELPHES 3 Collaboration],
  %``DELPHES 3, A modular framework for fast simulation of a generic collider
  %experiment,''
  JHEP {\bf 1402}, 057 (2014)
  [arXiv:1307.6346 [hep-ex]].
  %%CITATION = ARXIV:1307.6346;%%

\bibitem{Khachatryan:2015kon} 
  V.~Khachatryan {\it et al.} [CMS Collaboration],
  %``Search for Lepton-Flavour-Violating Decays of the Higgs Boson,''
  Phys.\ Lett.\ B {\bf 749}, 337 (2015)
  doi:10.1016/j.physletb.2015.07.053
  [arXiv:1502.07400 [hep-ex]].

\bibitem{hdecay}
A. Djouadi, J. Kalinowski, M. Spira
  [arXiv:9704448 [hep-ph]].
  %%CITATION = ARXIV:9704448;%%

\bibitem{atlas-tth-13}
%
%\cite{ATLAS:2016ldo}
%\bibitem{ATLAS:2016ldo}
  The ATLAS collaboration [ATLAS Collaboration],
  ``Search for the Associated Production of a Higgs Boson and a Top Quark
Pair in Multilepton Final States with the ATLAS Detector,''
  ATLAS-CONF-2016-058;
  %%CITATION = ATLAS-CONF-2016-058;%%
  %8 citations counted in INSPIRE as of 02 Jan 2017
%
  The ATLAS collaboration [ATLAS Collaboration],
``Measurement of fiducial, differential and production cross
sections in the $H\to\gamma\gamma$
decay channel with 13.3 fb-1 of 13 TeV proton-proton
collision data with the ATLAS detector",
ATLAS-CONF-2016-067;
%
  The ATLAS collaboration [ATLAS Collaboration],
``Search for the Standard Model Higgs boson produced in association with top
quarks and decaying into a bb pair in pp collisions at $\sqrt{s}$ = 13 TeV
with the ATLAS detector,'' ATLAS-CONF-2016-080;
%https://atlas.web.cern.ch/Atlas/GROUPS/PHYSICS/CONFNOTES/ATLAS-CONF-2016-080/;
%
%\bibitem{ATLAS:2016axz}
  The ATLAS collaboration [ATLAS Collaboration],
  ``Combination of the searches for Higgs boson production in association
with top quarks in the $\gamma\gamma$, multilepton, and $b\bar{b}$ decay
channels at $\sqrt{s}$=13 TeV with the ATLAS Detector,''
  ATLAS-CONF-2016-068.
  %%CITATION = ATLAS-CONF-2016-068;%%
  %5 citations counted in INSPIRE as of 02 Jan 2017

\bibitem{cms-tth-13}
%
%\cite{CMS:2016vqb}
%\bibitem{CMS:2016vqb}
  CMS Collaboration [CMS Collaboration],
  ``Search for associated production of Higgs bosons and top quarks in
multilepton final states at $\sqrt{s}=13~\mathrm{TeV}$,''
  CMS-PAS-HIG-16-022;
  %%CITATION = CMS-PAS-HIG-16-022;%%
  %11 citations counted in INSPIRE as of 02 Jan 2017
%
  CMS Collaboration [CMS Collaboration],
``Updated measurements of Higgs boson production in the
diphoton decay channel at $\sqrt{s}= 13$ TeV in pp collisions at
CMS", CMS-PAS-HIG-16-020.
%
%\bibitem{CMS:2016zbb}
  CMS Collaboration [CMS Collaboration],
  ``Search for $\mathrm{t\overline{t}H}$ production in the
$\mathrm{H}\rightarrow \mathrm{b\overline{b}}$ decay channel with
2016 pp collision data at $\sqrt{s}=13~\mathrm{TeV}$,''
  CMS-PAS-HIG-16-038.
  %%CITATION = CMS-PAS-HIG-16-038;%%
  %1 citations counted in INSPIRE as of 02 Jan 2017

\bibitem{thj-8}
CMS Collaboration [CMS Collaboration],
``Search for H to bbbar in association with single top quarks as a test of Higgs couplings,''
CMS-PAS-HIG-14-015;
C.~Boser [ATLAS and CMS Collaborations],
``Experimental searches for tHq,''
arXiv:1411.2988 [hep-ex];
A.~Popov [CMS Collaboration],
``Identification of signal events in a search for $H\to b\bar b$ produced in association with single top quarks,''
  arXiv:1411.7170 [hep-ex];
V.~Khachatryan {\it et al.} [CMS Collaboration],
``Search for the associated production of a Higgs boson with a single top quark in proton-proton collisions at $ \sqrt{s}=8 $ TeV,''
  JHEP {\bf 1606}, 177 (2016)
  doi:10.1007/JHEP06(2016)177
  [arXiv:1509.08159 [hep-ex]];
K.~Bloom,
``Search for associated production of a Higgs boson with a single top quark,''
  arXiv:1510.00894 [hep-ex];
CMS Collaboration [CMS Collaboration],
``Search for Associated Production of a Single Top Quark and a Higgs Boson in Leptonic Channels,''
  CMS-PAS-HIG-14-026;
L.~Caminada [CMS Collaboration],
``Higgs boson production in association with top quarks in CMS,''
  Nucl.\ Part.\ Phys.\ Proc.\  {\bf 270-272}, 217 (2016).
  doi:10.1016/j.nuclphysbps.2016.02.043     

\end{thebibliography}
\end{document}